\documentclass[runningheads]{llncs}

\usepackage[T1]{fontenc}
\usepackage{graphicx}
\usepackage{amsmath}
\usepackage{amssymb}
\usepackage{algorithm}
\usepackage{algpseudocode}
\usepackage{dirtytalk}
\usepackage{xcolor}
\usepackage{booktabs} 
\usepackage{caption}
\usepackage{subcaption}
\usepackage{multirow}
\usepackage{pgfplots}
\usepackage{hyperref}
\usepackage{adjustbox}
\usepackage{ulem}
\usepackage{verbatim}

\usepackage{epsfig}
\newcommand{\bigIT}[0]{\mathcal{I}} 
\newcommand{\smallIT}[1]{{\mathcal{I}}_{#1}} 
\newcommand{\dn}[1]{{\mathcal{D}}_{#1}} 

\title{Storing and Querying Evolving Graphs in NoSQL Storage Models}
\titlerunning{Storing and Querying Evolving Graphs in NoSQL Storage Models}

\author{Alexandros Spitalas\inst{1} \and Anastasios Gounaris\inst{2} \and Andreas Kosmatopoulos\inst{2} \and Kostas Tsichlas\inst{1}}

\authorrunning{A. Spitalas et al}

\institute{ University of Patras, Greece\\
            \email{\{ktsichlas,a.spitalas\}@ceid.upatras.gr}\\ \and
            Aristotle University of Thessaloniki, Greece\\
            \email{\{gounaria,akosmato\}@csd.auth.gr}}

\begin{document}

\maketitle

\begin{abstract}
This paper investigates advanced storage models for evolving graphs, focusing on the efficient management of historical data and the optimization of global query performance. Evolving graphs, which represent dynamic relationships between entities over time, present unique challenges in preserving their complete history while supporting complex analytical queries. We first do a fast review of the current state of the art focusing mainly on distributed historical graph databases to provide the context of our proposals. We investigate the implementation of an enhanced vertex-centric storage model in MongoDB that prioritizes space efficiency by leveraging in-database query mechanisms to minimize redundant data and reduce storage costs. 
To ensure broad applicability, we employ datasets, some of which are generated with the LDBC SNB generator, appropriately post-processed to utilize both snapshot- and interval-based representations. Our experimental results both in centralized and distributed infrastructures, demonstrate significant improvements in query performance, particularly for resource-intensive global queries that traditionally suffer from inefficiencies in entity-centric frameworks. The proposed model achieves these gains by optimizing memory usage, reducing client involvement, and exploiting the computational capabilities of MongoDB. By addressing key bottlenecks in the storage and processing of evolving graphs, this study demonstrates a step toward a robust and scalable framework for managing dynamic graph data. This work contributes to the growing field of temporal graph analytics by enabling more efficient exploration of historical data and facilitating real-time insights into the evolution of complex networks.
\end{abstract}

\section{Introduction} \label{sec:intro}

Consider the challenge of tracking the spread of an infectious disease through global air travel. To understand how an outbreak evolved, we need to analyze how passengers moved between cities over time, identifying critical points where the disease was likely transmitted. A static representation of the flight network is insufficient, as routes change daily, restrictions alter travel patterns, and the contagion follows a dynamic trajectory. Efficiently storing and querying this evolving network is essential to reconstruct past transmission pathways, predict future outbreaks, and inform containment strategies. The ability to retrieve historical snapshots at any time granularity and analyze evolving connectivity patterns is crucial for timely decision-making, highlighting the need for scalable and optimized temporal graph storage solutions.

The analysis of networks that evolve over time has gained increasing attention across multiple scientific disciplines, including sociology \cite{10.1086/428716}, physics \cite{doi:10.1142/S0219525911003050}, ecology \cite{rgd2018}, computer science \cite{DBLP:journals/cacm/MichailS18}, and engineering \cite{saidani2004}. These networks - commonly referred to as dynamic, adaptive, time-varying, evolving, or temporal networks, depending partly on the scientific area under consideration - capture interactions between entities that change over time. While different fields emphasize distinct aspects of temporal graphs, a common challenge remains: representing and querying the historical evolution of such networks efficiently. Understanding how relationships form, persist, and dissolve is critical for applications ranging from social influence modeling to infrastructure resilience analysis.

To provide a unified framework for studying these networks, the concept of Time-Varying Graphs (TVGs) \cite{doi:10.1080/17445760.2012.668546} has been introduced, offering a formalism to express various temporal network models. This formalism represents entities as nodes and their relationships as edges, both of which can be annotated with attributes (such as name and weight).  
TVGs facilitate the application of time-aware algorithms, enabling tasks such as temporal shortest paths, influence spread analysis, and anomaly detection in dynamic systems. However, while substantial progress has been made in defining the theoretical foundations and computational techniques for temporal graphs, less attention has been given to the underlying storage models that support these operations. Without efficient storage solutions, retrieving historical snapshots, executing time-dependent queries, and scaling to large datasets remain computationally expensive and impractical.
Indeed, many traditional database systems struggle to balance storage efficiency with query performance, especially for large-scale temporal graphs. This gap necessitates the development of storage models that can compactly represent historical data while enabling fast retrieval of both localized and global queries.  

To give an example, consider a graph corresponding to a social network that comprises millions of users (vertices) and the friendship relationships between them (edges). By examining the graph through two subsequent days, we witness a number of newly created, removed, and altered vertices or edges and a significant fraction of the graph that has remained the same across the two days. A system that aims to efficiently store all the snapshots of such a graph should employ techniques that mitigate the presence of unaltered data between different snapshots (i.e., take advantage of the commonalities between snapshots and refrain from storing duplicate data across snapshots). Moreover, the consideration of snapshots results in constraining the time granularity of the graph. Two consecutive snapshots could be defined over two consecutive years, months, days, hours, minutes, etc. This means that in general, snapshots are merely a view of the network at a particular time granularity and could change based on the user-defined query. However, in many systems, for performance reasons, the time granularity is predefined and cannot change since it requires the partial or total rebuilding of the system.

In a nutshell, this paper has a three-fold contribution; 1) it reviews the current research on historical graph management systems, 2) it evaluates extensively a vertex-centric approach that leverages MongoDB, and 3) it identifies a lack of temporal graph generators and makes a first step toward this direction by creating historical graph datasets based on the LDBC SNB \cite{angles2024ldbc}.\footnote{This paper partially contains heavily updated material from previously published conference papers by the same authors \cite{DBLP:conf/time/SpitalasGTK21,10.1007/978-3-031-33437-5_3}. In particular, apart from the new material explicitly stated in Sec. \ref{ssec:contr}, all experiments have been re-conducted in different settings (local and cluster modes).} 

The survey of existing storage techniques for temporal graphs, categorizes prior works based on their system architecture (distributed vs. non-distributed) focusing on distributed systems. In addition, it highlights trade-offs between different approaches and identifies key limitations in current solutions.

We also explore a vertex-centric storage model that leverages MongoDB to enhance the efficiency of temporal graph storage and retrieval. Unlike traditional snapshot-based approaches, this model optimizes space utilization by structuring historical data in a vertex-centric manner, reducing redundancy while maintaining fast access to both local and global queries. We compare our results to a similar Cassandra-based implementation \cite{DBLP:journals/computing/KosmatopoulosGT19}. Overall, the MongoDB implementation can perform more complex in-database queries and decrease client involvement in query processing. Additionally, instead of getting the documents from the database in a single large batch, we explore the option to employ a \texttt{foreach} approach (when this is expected to be more efficient), and as a result, to further reduce client involvement and memory requirements. 

Finally, to validate our approach we use/generate historical graph datasets and conduct extensive experiments to explore the performance of different versions of our storage model in various settings. These settings include transactions or graph-scale operations, batch or streaming modes, and the use of snapshots or time intervals. Our evaluation measures query execution times, storage efficiency, and scalability, with a focus on complex global queries that are traditionally expensive in entity-centric systems. The results both in a centralized as well as in a distributed setting, indicate significant improvements in execution speed, reduced memory consumption, and lower client-side processing overhead, allowing us to execute more expensive queries in the same infrastructure.
To generate some of the historical graph datasets we made some progress on creating a generator based on the LDBC SNB generator.

The structure of the paper is as follows. In Section~\ref{sec:review} we review the literature on distributed graph database systems that store the complete or the partial history of a historical graph. In Section~\ref{sec:background}, we describe the vertex-centric storage model given by some of the authors in \cite{DBLP:journals/dpd/KosmatopoulosTG17}, and provide details for implementation approaches of the vertex-centric schema in Cassandra as described in \cite{DBLP:journals/computing/KosmatopoulosGT19}. We describe our approach using MongoDB, which better exploits in-database query processing mechanisms in Section~\ref{sec:qp-new}. Our proposal is thoroughly evaluated in Section \ref{sec:exp} both in a centralized and in a distributed environment, and finally, we conclude in Section~\ref{sec:open}.

\section{Background and Related Work}  \label{sec:review}

There have been two main approaches regarding a TVG system's design \cite{Mondal:2012:MLD:2213836.2213854,deltagraph,kd15}, the time-centric approach and the entity-centric approach. In the former case, the system is indexed according to the time instants (i.e., changes are organized by the time instant they occur), while in the latter case, the system is indexed according to the entities, their relationships, and their respective history (i.e. changes are organized based on the vertex or edge they refer to). Most of the previous research work aims at storing the changes themselves (known as deltas) that occur between different snapshots. A system that maintains sets of deltas is thus able to reconstruct any particular snapshot by sequentially applying all the deltas up to the desired time instant. The above-described kind of framework can be used in both approaches but lends itself more naturally to a time-centric approach.

Entity-centric approaches can be separated into different groups of categories, with the most common being vertex-centric and edge-centric. In both approaches, the idea is to store the whole history of an entity (vertex or edge) in one entity, taking advantage of temporal locality and space efficiency by reducing duplicate information. In this case, for the reconstruction of a snapshot, access to every entity of the graph is required, which can be time-consuming. However, by taking advantage of locality and space efficiency, entity-centric approaches can reach if not surpass the efficiency of time-centric approaches. The advantage of entity-centric approaches is mostly seen in local and global queries when instead of looking at a single time instant (a single snapshot), the query involves a time interval (a series of snapshots).

Another viewpoint concerning a system's design is based on the type of queries that the system should be able to evaluate. Local queries involve a particular vertex or a limited selection of adjacent vertices (e.g., the $2$-hop neighborhood of a vertex). On the other hand, global queries consider the majority or the entirety of a graph's vertices (e.g., global clustering coefficient). Furthermore, both local and global queries should be able to be executed on either a time instant or on a range of time instants (e.g., average shortest path length between two vertices in the ten first snapshots). In the first case, a query aims to evaluate a measure at a particular time instant (e.g., shortest path length between two vertices at a particular time instant), while in the latter case, the query's objective is to extract information regarding a measure's evolution through snapshots for analytic purposes. By taking advantage of locality, entity-centric approaches will always be more efficient on local queries. However, regarding global queries, the efficiency depends on multiple factors, among which are the number of snapshots in the query and the specifics of the query. 

Due to the rapid growth of historical graphs, storage efficiency is of great importance. Time-centric approaches have to store logs of events instead of copying snapshots in every instant, to improve storage efficiency. However, depending on the occasion, different techniques can be more beneficial; for example, in the extreme case of changing completely all entities at every time instant, simply storing from scratch each snapshot is probably the best idea regarding space efficiency. Entity-centric approaches use logs of events within each entity. This is why, entity-centric approaches can be space-optimal and can probably reach a better combination of time and storage efficiency.

In the early approaches before 2017, there were two main research directions regarding evolving graph storage processing. Systems for non-evolving graphs, such as Trinity \cite{trinity}, could be leveraged to support historical queries by explicitly storing each snapshot, but apparently, such solutions are inefficient.
A comprehensive survey regarding these approaches for evolving graph data management can be found in \cite{algocloud} with the most notable proposals being those in \cite{deltagraph,gstar,gstaricde,ren11}. In general, these techniques rely on the storage of snapshots and deltas (logging), which exhibits a trade-off between space and time. Having a large number of snapshots results in deltas of small size but the space cost is substantial since we need to maintain many copies of the graph. On the other hand, having a handful of snapshots means that deltas will be quite large, and queries at specific time instants may require a long time to execute. 
Three of these proposals operate in a parallel or distributed setting, i.e., DeltaGraph \cite{deltagraph}, TGI \cite{kd15} and G* \cite{gstar}. The primary focus of DeltaGraph is the storage and retrieval of snapshots from an evolving graph sequence. It is best visualized as a rooted tree-like hierarchical graph structure with its leaves corresponding to snapshots of the sequence and the inner nodes corresponding to graphs that can be obtained by applying a differential function (e.g., intersection) to its children. DeltaGraph supports time point (single-point) queries, time interval snapshot queries, and multiple time point (multipoint) queries. Furthermore, along with the graph structure, a query is also able to return the attributes of vertices and edges (e.g., name or weight). The TGI system extends the functionality of the  DeltaGraph system by providing support for operations that are concerned with individual vertices or neighborhoods. 
The $G^*$ parallel system takes advantage of the commonalities that exist between snapshots by only storing each version of a vertex once and by avoiding storing redundant information that is not modified between different snapshots. Furthermore, $G^*$ achieves substantial data locality since each $G^*$ server is assigned its own set of vertices and corresponding entities. On the other hand, $G^*$ uses some form of logging to store connection information between different entities. 

Following the year 2017, a considerable amount of research has been conducted on historical graph management systems. A significant portion of these studies concentrates on enhancing the efficacy of main memory or streaming applications, as historical graphs tend to undergo frequent updates (e.g., \cite{10.1145/3364180,iyer2016time}). In contrast to earlier years, a greater number of studies now focus on entity-centric approaches and techniques that minimize the spatial complexity of the system. Furthermore, while many researchers employ an existing storage back-end to implement their approach, some leverage a customized storage back-end to optimize storage and indexing. Not all indexes can effectively support the queries frequently used in historical graphs, such as spatial or range queries.

It is worth noting that significant advancements have been made towards Online Transaction Processing (OLTP) and Online Analytical Processing (OLAP) systems for historical graph database systems. Many applications that involve historical graphs entail concurrent transaction processing, which requires careful attention. Additionally, it is imperative for such systems to not only store data but also execute a wide range of analytical tasks, including those specialized for historical graphs such as querying the graph's history. Despite these advancements, we have a long way to go before reaching a point where such systems will be able to efficiently handle both OLTP and OLAP queries since their requirements to achieve efficiency point toward different directions.

\subsection{Non-Distributed Systems}
In Table~\ref{tab:non-distr}, we provide a list of non-distributed systems for historical graph management. We do not elaborate on these systems since we focus on distributed systems. In Table~\ref{tab:distr}, we list all (to the best of our knowledge) distributed systems for historical graph management after the year $2017$ with a few earlier systems as well. Since we focus on large Historical Graphs, where storage efficiency is important, it makes sense that our setting falls under the distributed category, so we discuss briefly some of these systems. 

\begin{center}
\begin{table}
\resizebox{\textwidth}{!}{
\begin{tabular}{ |p{2.7cm}||p{3cm}|p{3cm}|p{5cm}|  }
 \hline
 \multicolumn{4}{|c|}{Summarizing the Characteristics of Non-Distributed Temporal Graph Management Systems} \\
 \hline
 \hline
 Systems & Memory & Storage Model & Time-related characteristics \\ [0.5ex] 
 \hline\hline
 InteractionGraph \cite{6702469} & Main Memory (old graph in disk) & Custom & Transaction time \\ 
 \hline
  STVG \cite{maduako2019stvg} & Main Memory & Neo4j & Valid time, offline, restricted to transit networks\\ 
  \hline
  ASPEN \cite{10.1145/3314221.3314598} & In-Memory/parallel & extends Ligra & Streaming \\ 
 \hline
  GraphOne \cite{10.1145/3364180} & In-memory NVMe SSD & Custom & Streaming, can't get arbitrary historic views if transaction time is assumed\\ 
 \hline
  Auxo \cite{8486795} & Main and External Memory & Custom & Transaction time \\ 
 \hline
 \cite{electronics9060895} & Main Memory & Custom & Transaction time, Snapshot-based, focus on space savings \\ 
 \hline
 \cite{andriamampianina:hal-03109670} & Main Memory  & Neo4j & Valid time, in addition to entity evolution it supports schema evolution\\ 
 \hline
  TGraph \cite{10.1145/2983323.2983335} & Main and External Memory & Neo4j & Support ACID Transactions, slow topological updates but fast property updates, Transaction time  \\ 
 \hline
  VersionTraveller \cite{ju2016version} & Main Memory & based on PowerGraph static graph management system & Offline Snapshot-based, focus on switching between snapshots \\ 
 \hline
 NVGraph \cite{8955949} & Non-Volatile Main Memory and DRAM & Custom & Online Snapshot-based, Transaction time \\ 
 \hline
 GraphVault \cite{iceis24} & Main and External Memory & LMDB (Lightning Memory-Mapped Database) & Uses time intervals with transaction time \\
 \hline
 AeonG \cite{10.14778/3648160.3648187} & Main and External Memory & Custom & Time intervals, Transaction time \\
 \hline
 Aion \cite{theodorakis2024aion} & Main and External Memory & Neo4J + Custom Indexes & Time intervals, Transaction time\\ [1ex]
 \hline
 
\end{tabular}}
\caption{Non-distributed systems for historical graph management. The second column (``Memory``) shows whether the system is implemented in main or external memory. The third column (``Storage Model``) shows whether they use a custom-based model or an already implemented one (like Neo4j). The third column (``Time-related characteristics``) shows the assumptions regarding the notion of time.} \label{tab:non-distr}
\end{table}
\end{center}

\subsection{Distributed Systems}
\paragraph{HiNode} This was the first storage model that adopted a pure vertex-centric approach. It was introduced in \cite{DBLP:journals/dpd/KosmatopoulosTG17} and supports valid time as well as extensions like multiple universes. It was implemented within the $G^*$ system \cite{gstar} by replacing its storage subsystem. They showed gains in space usage, which is an immediate consequence of the pure vertex-centric approach. They supported local queries (e.g., $2$-hop queries) as well as global queries (e.g., clustering coefficient). In addition, this vertex-centric model was also adapted for NoSQL databases by creating two models, SingleTable (ST) and MultipleTable (MT). In the former, all data fit in one table and a row has the data of a diachronic node, while in the latter, data are split into different tables (thus, slightly violating our pure vertex-centric approach). Two implementations were made, one in Cassandra \cite{DBLP:journals/computing/KosmatopoulosGT19} and later one in MongoDB \cite{DBLP:conf/time/SpitalasGTK21} for comparison reasons. In MongoDB, indices and iterative computation were used for efficiency and to reduce memory usage.

\begin{center}
\begin{table}
\resizebox{\textwidth}{!}{
\begin{tabular}{ |p{3cm}||p{3.2cm}|p{7cm}|  }
 \hline
 \multicolumn{3}{|c|}{Summarizing the Characteristics of Distributed Temporal Graph Management Systems} \\
 \hline
 \hline
 Systems & Storage Model & Time-related characteristics \\ [0.5ex] 
 \hline\hline
  Portal \cite{moffitt2017towards} & Spark's Dataframes & Offline, time as a property, valid time \\ 
 \hline
  GDBAlive \cite{mrpm20} & Cassandra & Transaction time \\ 
 \hline
 Graphsurge \cite{sahu2021graphsurge} & Custom & Offline snapshots, focus on differential computation across multiple snapshots \\ 
 \hline
  TEGRA \cite{265069} & Custom & Transaction time, based on persistent trees, incremental computation model, window analytics\\ 
 \hline
  GraphTau \cite{iyer2016time} & Spark's Dataframes & Streaming \\ 
 \hline
  Immortalgraph \cite{10.1145/2700302} & Custom & Transaction time, snapshot-based, focus on locality-aware (w.r.t. time and topology by replication), batch scheduling for computation\\ 
 \hline
  HGS \cite{DBLP:conf/edbt/KhuranaD16} & Cassandra & Transaction time, sophisticated based on snapshot \\ 
 \hline
 SystemG-MV \cite{8258092} & IBMs SystemG & Relaxed transaction time  \\ 
 \hline
  Raphtory \cite{STEER2020453} & Custom + Cassandra for archiving & Transaction time, streaming  \\ 
 \hline
  Chronograph \cite{8606161} & MongoDB & Offline, time as a property, focus on graph traversals\\ 
 \hline
 Graphite \cite{9101617} & Apache Giraph & Offline, application to time-dependent and time-independent algorithms \\ 
 \hline
  Granite \cite{RAMESH202194} & Based on Graphite & Focus on temporal path queries, partition techniques to keep everything in main memory \\ 
 \hline
  Tink \cite{lightenberg2018tink} & Apache Flink & Online, valid time \\ 
 \hline
  Gradoop - TPGM \cite{rost2019analyzing,christ2020distributed,Rost2021} & Apache HBase/ Accumulo & Valid and transaction time (bitemporal), fully-fledged system ranging from a graph analytical language to the storage model  \\ 
 \hline
  Greycat \cite{hfjrt17} & NoSQL Database + custom & Valid time, no edge attributes \\ 
 \hline
  PAST \cite{ding2019storing} & based on key/value stores (e.g., Cassandra) & Streaming spatio-temporal graphs, bipartite graphs, only edges with time-points, spatiotemporal-specific query workloads  \\ 
 \hline
 HINODE \cite{DBLP:journals/dpd/KosmatopoulosTG17,DBLP:journals/computing/KosmatopoulosGT19,DBLP:conf/time/SpitalasGTK21} & Custom (other versions are based on Cassandra and MongoDB) & Online, time as a property, valid time (allows more general notions of time), pure vertex-centric storage model \\
 \hline
\end{tabular}}
\caption{Distributed systems for historical graph management. The second column (``Storage Model``) shows whether they use a custom-based model or an already implemented one (like Neo4j). The third column (``Time-related characteristics``) shows the assumptions regarding the notion of time.}
\label{tab:distr}
\end{table}
\end{center}

\paragraph{Portal}
In \cite{moffitt2017towards,phdthesism}, the authors discuss interval-based models (where time is represented by intervals) and point-based models (where time is represented by a sequence of time instants) for time queries, focusing on the interval-based model with sequenced semantics. They propose a Temporal Graph Algebra (TGA) and a temporal graph model (TGraph) supporting TGA.
In addition, in \cite{DBLP:journals/corr/MoffittS16} they propose a declarative language (Portal) based on the previous model and built on top of a distributed system (Apache Spark). The Portal has SQL-like syntax that follows the SQL:2011 standard. They also discuss possible algorithms on temporal graphs among which are node influence over time, graph centrality over time, communities over time, and spread of information. TGraph is a valid time model that extends the property graph model (each edge and vertex is associated with a period of validity), while all relations must meet 5 criteria: uniqueness of vertices/edges, referential integrity, coalesced, required property, and constant edge association. TGA is both snapshot and extended snapshot reducible presenting a new primitive (resolve) while containing operators like trim, map, and aggregation. Portal uses Spark for in-memory representation and processing while it uses Apache Parquet for on-disk data layout using node files and edge files (but it doesn't support an index mechanism). They experimented with different in-memory representations: a) SnapshotGraph (SG), which stores the graph as individual snapshots, b) MultiGraph (MG), which stores one single graph by storing one vertex for all periods and one edge for every time period and c) OneGraph that stores each edge and vertex only once. It has distributed locality like Immortalgraph \cite{10.1145/2700302}, and they experimented with different partitioning methods (equi-depth partitioning is more efficient in most experiments). They store materialized nodes/edges instead of deltas and they also experimented with both structural and temporal locality, concluding that temporal locality is more efficient, among other reasons, due to the lack of sufficient discrimination in the temporal ranges of the datasets.

\paragraph{ImmortalGraph}
\cite{10.1145/2700302} is a parallel in-memory storage and computation system for multicore machines in a distributed setting designed for historical graphs. It focuses on locality optimizations, both in the storage of the data and in the execution of the queries using locality-aware batch scheduling (LABS). 
It supports parallel temporal graph mining using iterative computations. ImmortalGraph supports both global and local queries at a point in time or a time window. Data are stored in snapshot groups using either edge files or vertex files, depending on the application. A snapshot group organizes together snapshots of a time interval by storing the first snapshot and then logging the rest of the changes. Either time locality can be ensured by grouping activities associated with a vertex (and a vertex index) or structural locality can be ensured by storing together neighboring vertices (and a time index). In order to combine the advantages of both approaches, they replicate the needed data and decide which approach to use according to the type of query and the distance from the snapshot of the group. LABS favors partition-parallelism from snapshot-parallelism, so they prefer batch operations of vertex/edges achieving better locality and less inter-core communication. They also experimented with iterative graph mining and iterative computations. For the former, they reconstruct the needed snapshots in memory favoring time locality (and they compare both push, pull, and stream techniques), while for the latter they compute the first snapshot and the succeeding snapshots in batch (achieving better locality). They also implemented both low-level and high-level query interfaces, the latter used for iterative computations. An earlier implementation of ImmortalGraph is Chronos \cite{10.1145/2592798.2592799} with the main difference being that it only focuses on time locality. Finally, they provide a low-level as well as a high-level programming interface (APIs), which can be considered as an analytics engine. They experiment on Pagerank, diameter, SSSP, connected components, maximal independent sets, and sparse-matrix vector multiplication.

\paragraph{Historical Graph Store (HGS)}
\cite{DBLP:conf/edbt/KhuranaD16} is a cloud parallel node-centric distributed system for managing and analyzing historical graphs. HGS consists of two major components, the Temporal Graph Index (TGI), which manages the storage of the graph in a distributed Cassandra environment, and the Temporal Graph Analysis Framework (TAF), which is a Spark-based library for analyzing the graph in a cluster environment. TGI combines partitioned eventlists, which store atomic changes, with derived partitioned snapshots, which is a tree structure where each parent is the intersection of children deltas (used for better structure locality storing neighborhoods).  Both of them are partitioned, while they are also combined with a version chain to maintain pointers to all references of nodes in chronological order. TGI divides the graph into time spans (like the snapshot groups of ImmortalGraph) with micro-deltas, which are stored as key-value pairs contiguously into horizontal partitions at every time span. In this way, it can execute in parallel every query in many query processors and aggregate the result to the query manager or to the client. It can work both on hash-based and locality-aware partitioning by projecting a time range (time-span) of the graph in a static graph. TAF supports both point-in-time queries and time-window queries. Some of the supported queries are subgraph retrieval with filtering, aggregation, pattern matching, and queries about the evolution of the graph. An earlier implementation of TGI is DeltaGraph \cite{6544892}, which focuses on snapshot retrieval.

\paragraph{ChronoGraph}
\cite{8606161} is a temporal property graph database built by extending Tinkerpop and its graph traversal language Gremlin so as to support temporal queries. It stores the temporal graph in persistent storage (MongoDB), and then loads the graph in-memory and traverses it. Their innovation is not in the storage model but in how they support traversal queries efficiently on top of it. It exploits parallelism, the temporal support of Tinkerpop to increase efficiency, and lazy evaluations to reduce memory footprints of traversals. Its main focus is on temporal graph traversals but can also return snapshots of the graph. They distinguish point-based events and period-based events because of their semantics and their architectural needs. They use aggregation to convert point-based events to period-based events so as not to have two different semantics in order to improve time efficiency in query execution.
They achieve this by using a threshold as the maximum time interval that may exist between time instants so as to group them together. A graph is composed of a static graph, a time-instant property graph, and a time-period property graph. They also use event logic, where an event might be either a vertex or an edge, during a period of time or a time instant. They implemented temporal versions of BFS, SSSP, and DFS. However, they don't recommend DFS on their system because of Gremlin's recursive logic. An extension of Chronograph by using time-centric computation for traversals is given in \cite{9128046}.

\paragraph{Tink}
 \cite{lightenberg2018tink} is an open-source parallel distributed temporal graph analytics library built on top of the Dataset API of Apache Flink and uses the programming language Gelly. It extends the temporal property graph-model focusing on keeping intervals instead of time-points by saving nodes as tuples. It depends on Flink to use parallelism, optimizations, fault tolerance, and lazy-loading and supports iterative processing. It also uses functions from Flink like filtering, mapping, joining, and grouping. Most algorithms use Gelly's Signal/Collect (scatter-gather) model, which executes computations in a vertex-centric way. It also provides temporal analytic metrics and algorithms. For the latter, they implemented shortest path earliest arrival time and shortest path fastest path while for temporal metrics they provide temporal betweenness and temporal closeness.

\paragraph{Gradoop (TPGM)}
Temporal Property Graph Model (TPGM) \cite{rost2019analyzing,christ2020distributed,Rost2021}  is an extension of Gradoop's Extended Property Graph Model - EPGM - (model for static graph processing, presented in a series of papers from 2015, see \cite{10.1145/2980523.2980527}) to support temporal analytics on evolving property graphs (or collection of graphs) that can be used through Java API or with KNIME. Gradoop is an open-source parallel distributed dataflow framework that runs on shared-nothing clusters and uses GRALA as a declarative analytical language and TemporalGDL as a query language. Gradoop supports Apache HBase, and Apache Accumulo to provide storage capabilities on top of HDFS, while other databases can also be used with some extra work. TPGM supports bitemporal time by maintaining logical attributes for start and end time for both valid and transaction time for every graph entity. While TPGM provides an abstraction, Apache Flink is used for handling the execution process in a lazy way. GRALA provides operators both for single graphs and graph collections, it supports retrieval of snapshots, transformations of attributes or properties, subgraph extraction, the difference of two snapshots, time-dependent graph grouping, temporal pattern matching, and others. For some more complex algorithms, it also supports iterative execution using Apache Flink's Gelly library. 
Lastly, they have implemented in Flink a set of operations for their analytics engine - by using Flink Gelly. They provide an extensive description of their architecture, while they also provide a \textit{Lessons Learned} section that contains valuable information concerning their design choices.

\paragraph{SystemG-MV}
In \cite{8258092} they propose an OLTP-oriented distributed temporal property graph database (dynamically evolving temporal graphs). It is built on top of IBM's SystemG, which is a distributed graph database using LMDB ($B$-tree-based key-value store). Data are stored in tables with key/value pairs allowing to query part of the graph efficiently without retrieving the whole snapshots. Different tables exist for vertices, edges, and properties, while it supports updates only on present/future timestamps like transaction-time models. Therefore, changing previous values of the graph is not allowed explicitly, but it is possible to change past events by using low-level methods. In this model, they save two timestamps for the creation/deletion of vertices/edges but they don't allow edges to be recreated with the same id, although multiple edges can exist between a pair of vertices. For vertices, they keep the deleted vertices in a different table, while for properties they keep only one timestamp per update, as the rest can be calculated. Alongside the historic tables, they keep one table with the current state of the graph for more efficient queries. Their low-level algorithms that constitute the API of their storage model resemble the API suggested in \cite{DBLP:journals/dpd/KosmatopoulosTG17}.

\paragraph{TEGRA}
\cite{265069} is a distributed system with a compact in-memory representation (using their custom storage model) both for graph and intermediate computation state. Its main focus is on time window analytics for historical graphs, but it can also be used for live analytics as the data are ingested in the database. An interesting feature is the ICE computational model that takes advantage of the intermediate state of computations by storing it and using it in the same or similar queries. Computations are performed only on subgraphs affected by updates at each iteration. This has some overhead related to finding the correct state and the extra entities that should be included in the query when there are many updates at each iteration or while trying to use ICE on different queries. Tegra also uses TimeLapse, an API for high-level abstraction that allows what-if questions that change the graph creating different histories, suited for data analytics purposes. 
The storage model behind TEGRA is DGSI, which uses persistent data structures to maintain previous versions of data during updates. In particular, it uses persistent Adaptive Radix Trees (pART) to store edges and nodes separately by employing the path-copying technique \cite{DRISCOLL198986}. It uses simple partitioning strategies to distribute the graph to nodes. Log files are used to store updates between snapshots, which are stored in turn in the two pARTs. They use branch-and-commit primitives in tandem with the GAS (Gather - Apply - Scatter) model \cite{10.5555/2387880.2387883}. It also supports changing any version, thus leading to a branched history like a tree (reminiscent of full persistence \cite{DRISCOLL198986}). Lastly, TEGRA also uses an LRU policy to periodically remove versions that have not been accessed for a long time.

\paragraph{Graphite}
\cite{9101617} is a distributed system for managing historical graphs (offline with valid time) by using an interval-centric computing model (ICM) built over Apache Giraph. They assume data are given in ascending time order and any vertex can exist only once for a contiguous time interval. It can execute both time-independent and time-dependent historical queries (temporal queries on a time window). They introduce a unique time-warp operator for temporal partitioning and grouping of messages that hides the complexity of designing temporal algorithms while avoiding redundancy in user logic calls and messages sent. 
ICM uses Bulk Synchronous Parallel (BSP) execution for every active vertex of a query until it converges. They use two stages of logic, compute and scatter, where compute does the computations needed for a vertex, and scatter transfers it with messages to neighbor vertices as needed. The time-warp operator is applied at the alternating compute scatter steps to help the sharing of calls and messages across intervals. A key aspect is that it correctly groups the input with no duplicates, while it returns the minimum possible number of triples. They also designed and constructed a plethora of time-independent and time-dependent algorithms for their system with an extensive experimental evaluation.

\paragraph{Granite}
\cite{RAMESH202194} is a distributed engine for storing and analyzing temporal property graphs (supports temporal path queries) made on top of and as a sequel to Graphite focusing on path queries. Its design is based on the assumption that the workloads consist of infrequent updates and frequent queries. They extend the previous model by adding a temporal aggregation operator, indexing, query planning, and optimization, while they prefer to relax ICM so as to make it work beyond time-respecting algorithms. 
Granite handles both static temporal graphs and dynamic temporal graphs while it uses interval-centric features only in the latter. To optimize path queries, they split them and execute them concurrently, while they also keep statistics about the active nodes at each time point so as to optimize the query planning. While Graphite makes hash partitioning at query execution, Granite first partitions every entity according to its type, and then it performs a topological partition to its independent group of entities of the same type and splits them into workers using the round-robin technique. They also use a result tree so as not to send duplicate paths across the system (some parts of the path might be the same).
Lastly, they propose a query language for path queries.

\section{Preliminaries} \label{sec:background}

Let $G = (V_T,E_T)$ be a static historical network. The set of historical nodes $V_T$ consists of a set of nodes along with their time intervals, that is $V_T\subset V\times\mathbb{N}^2$. The set of historical edges $E_T$ is a set of edges along with their time intervals, that is $E_T\subset E\times \mathbb{N}^2$, where $E$ contains all possible $|V| \choose 2$ undirected edges. Note that we consider nodes and edges that have a single valid time interval, but it is easy to generalize to a set of valid time intervals. The state of the graph $G$ at a particular time instant $t$, is called a snapshot and it is denoted by $G_t$. The preceding definitions mean that each node $v\in V$ (and edge $e\in E$) has a time interval attached $[t^{(s)}_v,t^{(f)}_v]$ (similarly $[t^{(s)}_e,t^{(f)}_e]$) (where $(s)$ and $(f)$ stand for start and finish respectively) that dictates the time instants where node $v$ (edge $e$) is existent. Thus, if $t\not \in [t^{(s)}_v,t^{(f)}_v]$, then $v$ is not existent at time $t$. $V_{ij}\subseteq V$ contains all nodes that have a time interval that spans the query interval $[t_i,t_j]$. The time interval of each edge is by definition a subset of the time interval of the respective vertices. 
In case of multiple time intervals, we have to define the borders of each interval accordingly, to avoid overlaps. For example, each interval should be open at the left and closed at the right. The convention we make is that a time point $t$ is represented by $(t,t]$.
Assume that by $\mathcal{N}_{ij}(v)$ we represent the neighborhood of node $v$ in the query time interval $[t_i,t_j]$. Note that $\mathcal{N}_{ij}(v)$ may even be the empty set for specific query time intervals. 

\subsection{HiNode Theoretical Model}
In~\cite{DBLP:journals/dpd/KosmatopoulosTG17}, the first purely entity-centric, and more specifically, vertex-centric model for maintaining graph historical data, termed HiNode is introduced. Its strongest point is that it builds upon a theoretical storage model that is asymptotically space-optimal. The core idea behind HiNode's solution is that a vertex history throughout all snapshots is combined into a set of collections called diachronic node. The diachronic node utilizes indexes to model a vertex's history as a collection of intervals that permit geometric operations upon them. 
In particular, a vertex $v\in G_i$ is characterized by a set of attributes (e.g., color), a set of incoming edges from the other vertices of $G_i$, and a set of outgoing edges to the other vertices of $G_i$. We construct an external interval tree $\bigIT$ that maintains a set of intervals $[t_v^{(s)},t_v^{(f)}]$ for each vertex $v$. We mark a vertex to be ``active'' (alive) up until the latest time instant, by setting the $t^{(f)}$ value to be $+\infty$. 
Finally, each interval $[t_v^{(s)},t_v^{(f)}]$ is augmented with a pointer (handle) to an additional data structure for each vertex $v$, corresponding to the diachronic node.

A diachronic node $\dn{v}$ of a vertex $v$ maintains a collection of data structures corresponding to the full vertex history of $G$, i.e., when that vertex was inserted, all corresponding changes to its edges or attributes and finally its deletion time (if applicable). More formally, a diachronic node $\dn{v}$ maintains an external interval tree $\smallIT{v}$ that stores information regarding all of $v$'s characteristics (attributes and edges) throughout the entire history. An interval in $\smallIT{v}$ is stored as a quadruple $(f,\{\ell_1, \ell_2, \ldots\},t^{(s)},t^{(f)})$, where $f$ is the identifier of the attribute that has values $\ell_1$, $\ell_2$, $\ldots$ during the time interval $[t^{(s)},t^{(f)}]$. Note that an edge of $v$ (i.e., one endpoint of the edge is $v$), can be represented as an attribute of $v$ by using one value $\ell_i$ to denote the other end of the edge, another value $\ell_j$ to mark the edge as incoming or outgoing and a last value $\ell_h$ that is used as a handle to the diachronic node corresponding to the vertex in the other end of the edge. The remaining $\ell$ values can be used to store the attributes of the edge themselves (e.g., weight).
Additionally, the diachronic node maintains a $B$-tree for each attribute and for each individual edge of the vertex. Full details are in \cite{DBLP:journals/dpd/KosmatopoulosTG17}.

HiNode supports adding or removing vertices and attributes as fundamental operations upon which more complex operations and queries (e.g., graph traversal, shortest path evaluation, etc.) are constructed. In HiNode, each change is stored $O(1)$ times, resulting in an asymptotically optimal total space cost. Furthermore, due to the local handling of history, HiNode performs well on local queries and the authors further demonstrate that HiNode on top of $G^*$ is competitive regarding global queries as well compared to $G^*$ \cite{DBLP:journals/dpd/KosmatopoulosTG17}.

\subsection{Implementation in Cassandra} \label{sec:current} 

The first HiNode implementation, hereafter termed as HiNode-$G^*$\footnote{Source code available at \url{https://github.com/hinodeauthors/hinode}}, was based on extensions to the $G^*$ ~\cite{gstar,gstaricde} distributed graph database. This design choice incurred severe limitations regarding the efficiency and scalability of the HiNode-$G^*$ prototype. In a follow-up work \cite{DBLP:journals/computing/KosmatopoulosGT19}, to outperform solutions based on tailored graph management systems, such as Neo4j, we proposed to leverage NoSQL as the underlying database technology providing preliminary results with respect to two different implementation approaches in Cassandra. These approaches consist of the Single Table (ST) and Multiple Table (MT) implementations. In the former case, the entire history of a vertex is stored in a single table with each vertex corresponding to a single table row, while in the latter case, the data of each vertex is stored in multiple tables with each table corresponding to a single vertex attribute.  \footnote{Source code available at \url{https://github.com/akosmato/HinodeNoSQL}}

To adequately support global queries (i.e., queries that involve a significant part of a snapshot's vertices), the two models offer two querying modes for the retrieval of all vertices relevant to a specified query.
Let $[t_s,t_e]$ be a specified time range for which a query is about to be executed. In the first mode (termed Retrieve\_All (RA)), and regardless of the given time range, we retrieve all vertices from each model and then perform a client-side filtering operation, where we discard any vertices that do not belong in $[t_s,t_e]$.
In the second mode (termed Retrieve\_Relevant (RR)), in each model, we first determine the vertices that are ``alive'' at $[t_s,t_e]$, and then, we retrieve them. 

While in ST the implementation of RR is straightforward, MT requires additional work since retrieving a particular (set of) attribute(s) during a certain time interval $[t_s, t_e]$ would translate to a range query and the retrieval of all data with a ``timestamp'' value between $t_s$ and $t_e$ (i.e., we are not interested in any updates that occur outside $[t_s, t_e]$). Since Cassandra does not natively permit double-bounded range queries for the sake of efficiency, we fetch the relevant data with a timestamp larger than $t_s$ and then filter all data with a timestamp larger than $t_e$ at the client side. In~\cite{DBLP:journals/computing/KosmatopoulosGT19} there is extensive experimental evaluation, with interesting outcomes which are not presented here due to space limitations.
All in all, the choice of a particular vertex-centric implementation is not straightforward and exhibits different trade-offs depending on the query at hand.

\section{The MongoDB Implementation}
\label{sec:qp-new}
Our main motivation behind using MongoDB is to exploit the wider range of indexing options and its capabilities to reduce client involvement when processing queries. Additionally, in Cassandra, data are saved as strings and, as such, they are serialized when returned to the client, while in MongoDB we can store the elements of the nodes with a combination of lists and documents. Overall, we can perform more complex in-database queries and decrease client involvement in query processing. Finally, in the new implementation, instead of getting the documents from the database in a single large batch, we have the option to employ a \texttt{foreach} approach (when this is expected to be more efficient) and as a result, to mitigate intermediate client-side storage requirements\footnote{Source code available at \url{https://github.com/alexspitalas/HiNode-MongoDB/}}.

\subsection{Schema Alternatives} \label{ssec:schema_alts}

Both the ST and MT models have been transformed to comply with MongoDB's JSON format in a straightforward manner. In addition, we developed an alternative schema for both models, where the elements of the primary key are inserted as characteristics in the document; the primary key is the standard key assigned automatically by MongoDB. The reason for this schema is to further simplify the client-side tasks (i.e., the processing refers to the document content exclusively) with no difference in the capability of answering specific types of queries.

In the ST model, a document is a representation of a diachronic node and consists of the primary key as a triple (\texttt{vid, start} and \texttt{end} of the node), the incoming and outgoing edges, and the vertex attributes. The features forming the key are stored as atomic string values, while the vertex attributes are stored as a list of sub-documents, where each document is a triple. 
The incoming and outgoing edge metadata are stored as a sub-document containing a list of triples (where each triple is a MongoDB sub-document). The former document is essentially a hashmap structure with the key corresponding to the vertex id, while the nested sub-document stores the attributes and the period for each edge.
The following three indices are built: (i) an index on \texttt{vid}; (ii) an index on \texttt{start} and \texttt{end}; (iii) an index on the complete key. The first index
allows quick retrieval of a specific vertex, while the second and third indices facilitate stabbing queries.

In the MT model, we split the diachronic node into three sets of collections, one containing vertices, one containing incoming edges, and one containing outgoing edges. Each set consists of one collection concerning the existence period of the vertex or edge and one collection for every attribute. 
The standard indices are on \texttt{vid}, \texttt{(vid,start)} and \texttt{(vid,start,end)} in the first set of collections.  For the edge collection sets, the multikey indices are on \texttt{sourceID} (or \texttt{targetID}) and the \texttt{start} timestamp. 

In summary, the main difference with the Cassandra-based implementation in \cite{DBLP:journals/computing/KosmatopoulosGT19} in terms of modelling is the increased flexibility regarding indices and the fact that sub-documents are stored without being serialized as strings, thus making able to run more complex queries.

\subsection{Query processing} \label{ssec:query_pro}

For local queries, the server (database) side is straightforward, while most of the work is performed on the client side. The local queries we investigate in this paper are the retrieval of the history of a vertex and one-hop queries. In the former query, we must retrieve the history of a specified vertex for a time interval. In the latter query, we must retrieve the neighbors of a vertex at a specified time interval. Both queries are straightforwardly supported by both implementation models.

Because of the vertex-centric approach, global queries are more meaningful to investigate in depth, aiming at rendering them more efficient. Global query processing is comprised of two phases. The first phase is related to the retrieval of the data, while the second phase is related to the processing of the retrieved data. These phases can be intertwined. In our implementation, the two phases are separated so that the client's side is the same for all ST-based and all MT-based approaches, respectively. Regarding the retrieval of the data, three variants have been developed, Retrieve\_Relevant~(RR), Retrieve\_All~(RA), and In-Database~(ID).

In RR, the main objective is to find the relevant documents by retrieving only their necessary characteristics.  In the RA approach, we retrieve all the characteristics of the document, while we check if the document is needed for the query. In ST model, a document consists of all information of a diachronic node (edges and properties), while in MT model, a document is referring to either a node, an edge or a property and the corresponding information in each case. Comparing this method with RR, we perform only one read at the database, but we retrieve more data than necessary if the document is not needed for the query; as a result, RR is expected to perform better when the amount of data stored per node is much higher than those needed to establish if the node is relevant to the query. This relevance check, along with the rest of the query execution, is performed on the client. In the new MongoDB implementation, contrary to the initial implementation based on Cassandra, we adopt a more incremental (iterative) approach instead of returning all data in a single batch; this has increased the scalability of global queries so that they can be executed without throwing an out-of-memory error.

However, the most notable difference between the two implementations is that MongoDB naturally lends itself to in-database query processing, so that the client gets only the data needed to compute the final results and not a superset of these data (through submitting more complex queries as supported by the MongoDB driver). To this end, we use the in-database MongoDB mechanisms to perform the relevance checks mentioned in the RR approach.
Similarly to RR, the data needed for the final answer computations are returned incrementally to the client. As such, this approach has even lower space requirements on the client side, and at the same time, it allows for both the server and the client side to work in parallel\footnote{In some local queries (like one-hop query), it may make sense to adopt an in-database query processing rationale, but this is beyond the scope of this paper}.

\subsection{Transactions with MongoDB} \label{ssec:transaction_mdb}

In a distributed environment, supporting both OLTP and OLAP queries in the same database is challenging, although some efforts have been made in this area. In RDBMS databases, OLTP queries are supported by ensuring ACID properties. However, these properties are more difficult to maintain in a distributed environment, especially in NoSQL databases. In our case, the specifics of the OLTP and OLAP queries we aim to support are summarized below.

\begin{itemize}
\item Support of OLTP queries: These are simple queries typically involving a few records. The emphasis is on fast processing, as OLTP databases are frequently read, written, and updated. If a transaction fails, the built-in system logic ensures data integrity.
\item Support of OLAP queries: These are complex queries usually referring to a large part of the graph. The emphasis here is on efficiency in executing the queries without concern for parallel updates performed on the graph.
\end{itemize}

To address these requirements, MongoDB supports multi-document ACID transactions with minimal additional code. Due to multi-document ACID support, if a transaction involves multiple documents, it will maintain Atomicity, Consistency, Isolation, and Durability throughout the operation, leaving the database in a consistent state. This feature allows us to support transactions in our MT model with MongoDB. To the best of our knowledge, some NoSQL databases only support ACID properties up to one document. In contrast, many NoSQL databases do not support ACID transactions at all, instead adopting a BASE transaction model that prioritizes availability over consistency.

\section{Experiments} \label{sec:exp}
Experiments were carried out both locally and in a cluster, with the aim of obtaining different conclusions in each setting. The single node system has an AMD Ryzen 9 7900X CPU @ 4.70GHz, 196GB DDR5 RAM, and a 1TB NVMe, while the client and the databases are co-located on the same machine. The client application was written in Java.

For the cluster experiments, we used a $4$-node cluster, where all the machines of the cluster have the following specifications: Intel(R) Core(TM) i7-10700 CPU @ 2.90GHz, 16GB RAM, and a 1TB NVMe disk. The client and the database are co-located on the same network with a 10GBps LAN connection but in a different machine. The client machine is equipped with an Intel(R) Core(TM) i3-7100 CPU @ 3.90GHz, 4GB RAM, and a 500GB HDD.

For the experiments, we employ two queries, one local and one global query: 
1) \textit{one-hop} (local query): that returns all nodes that are adjacent to a given node within a time interval, 
2) \textit{vertex degree distribution} (global query): that returns the degree distribution of a graph within a time interval. These queries are applied on three different datasets that are shown in Table~\ref{tab:datasets}.
We experimented with all different combinations of the MT and ST models, and the Cassandra and MongoDB systems. For MongoDB, we test with all modes of global query processing (retrieve-all, retrieve-relevant, in-database). Each query refers to a range of snapshots from $1$ to $all$.

\subsection{Experiments on Snapshot-based Datasets} \label{sec:exp-snap}

For the snapshot-based datasets, the temporal domain is discretized into a finite set of discrete time instants within which, all time-relative entities are aligned. This approach offers increased query efficiency and facilitates conversion of the dynamic graph into a static one at a specific time-instant. However, it is not always feasible to represent the problem using snapshot-based datasets due to certain limitations.

\begin{table}[tb!]
\begin{center}
\begin{tabular}{|c|c|c|c|}
\hline
 Name & \# of vertices & \# of edges & \# of snapshots  \\ \hline
 hep-Th \cite{snap_cit_hepth} & 27.77K & 352.8K & 156  \\ \hline
 hep-Ph \cite{snap_cit_hepph} & 34.5K & 421.6K & 132  \\ \hline
 US Patents \cite{snap_cit_patents} & 3774.8K & 16.5M & 444  \\ \hline
 \end{tabular}
\caption{The datasets used in the snapshot experiments.}
\label{tab:datasets}
\end{center}
\end{table}

\subsubsection{Local Queries} \label{ssec:local_queries}

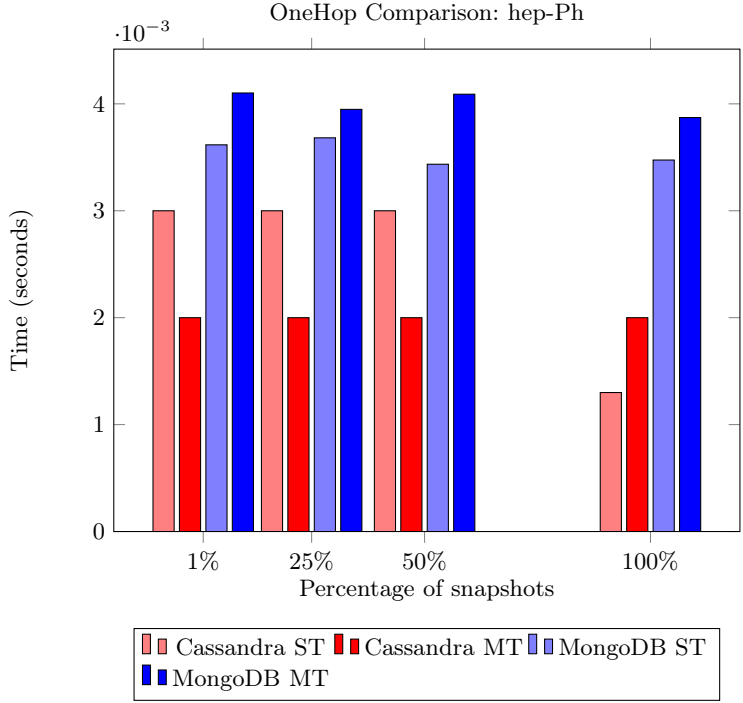
\begin{figure}[tb!]
    \centering
    \begin{tikzpicture}
    \begin{axis}[
        ybar,
        bar width=8pt,
        width=0.65\textwidth,
        height=8cm,
        xlabel={Percentage of snapshots},
        ylabel={Time (seconds)},
        xtick={1, 25, 50, 100},
        xticklabels={1\%, 25\%, 50\%, 100\%},
        ymin=0,
        enlarge x limits=0.2,
        legend style={at={(0.5,-0.2)}, anchor=north, legend columns=3},
        title={OneHop Comparison: hep-Ph}
    ]
    \addplot[fill=red!50] coordinates {(1, 0.003) (25, 0.003) (50, 0.003) (100, 0.0013)};
    \addplot[fill=red!100] coordinates {(1, 0.002) (25, 0.002) (50, 0.002) (100, 0.002)};
    \addplot[fill=blue!50] coordinates {(1, 0.003616784) (25, 0.003682238) (50, 0.003435285) (100, 0.003474553)};
    \addplot[fill=blue!100] coordinates {(1, 0.004101663) (25, 0.003948437) (50, 0.004090241) (100, 0.003871774)};
    
    \legend{Cassandra ST, Cassandra MT, MongoDB ST, MongoDB MT}
    \end{axis}
    \end{tikzpicture}
    \caption{OneHop query comparison for the hep-Ph dataset in the Cluster.}\label{cl::onehopPh}
\end{figure}

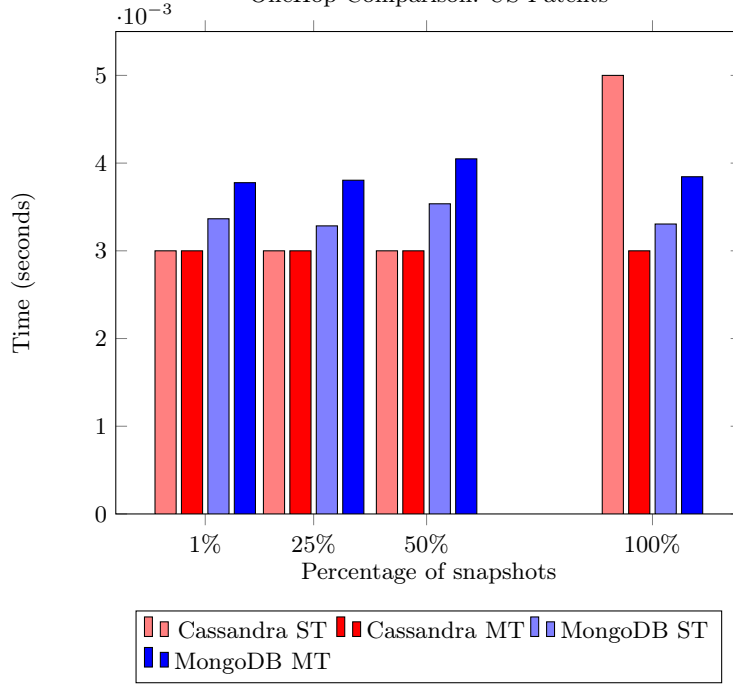
\begin{figure}[tb!]
    \centering
    \begin{tikzpicture}
    \begin{axis}[
        ybar,
        bar width=8pt,
        width=0.65\textwidth,
        height=8cm,
        xlabel={Percentage of snapshots},
        ylabel={Time (seconds)},
        xtick={1, 25, 50, 100},
        xticklabels={1\%, 25\%, 50\%, 100\%},
        ymin=0,
        enlarge x limits=0.2,
        legend style={at={(0.5,-0.2)}, anchor=north, legend columns=3},
        title={OneHop Comparison: US Patents}
    ]
    \addplot[fill=red!50] coordinates{(1, 0.003) (25, 0.003) (50, 0.003) (100, 0.005)};
    \addplot[fill=red!100] coordinates{(1, 0.003) (25, 0.003) (50, 0.003) (100, 0.003)};
    \addplot[fill=blue!50]  coordinates{(1, 0.003365316) (25, 0.003284298) (50, 0.003536117) (100, 0.003305547)};
    \addplot[fill=blue!100]  coordinates{(1, 0.003776864) (25, 0.003804872) (50, 0.004048564) (100, 0.003844557)};
    
    \legend{Cassandra ST, Cassandra MT, MongoDB ST, MongoDB MT}
    \end{axis}
    \end{tikzpicture}
    \caption{OneHop query comparison for the US Patents dataset in the Cluster.}\label{cl::onehopus}
\end{figure}

Regarding local queries on a cluster environment, we executed one-hop queries for all three snapshot-based datasets, retrieving data from $1$ snapshot, $25\%$ of the snapshots, $50\%$ of the snapshots and $100\%$, so as to observe the scaling of the models while we retrieve more historical data. We repeated each query $500$ times and we report the average values. For each set of $500$ runs we have a cold start. In Figures \ref{cl::onehopPh}, and \ref{cl::onehopus} we observe the performance of every model in each dataset, reaching the following conclusions (Hep-Th has similar results with Hep-Ph and it is not depicted):

\begin{enumerate}
    \item The most important observation is that in general, Cassandra MT outperforms Cassandra ST, managing to decrease time up to $84.6\%$, (the gap decreases as we move to bigger datasets). However, MongoDB ST slightly outperforms MongoDB MT, managing to decrease time by $6.7\%$ up to $24.9\%$. This difference can be attributed to the fact that MongoDB is a document store, and as such, it can better cope with complex collections like in the case of ST, where all data of a node are placed together.
    
    \item Another observation is that in MongoDB, the performance is stable as we increase the percentage of snapshots used in the query. This is expected in a local query at a vertex-centric system, while in Cassandra there are some spikes when we use $100\%$ of the snapshots.

    \item While querying for up to $50\%$ of the snapshots, Cassandra MT is the best overall model, managing to decrease the running time between $8.6\%$ up to $45.6\%$. However, when the query involves $100\%$ of the snapshots, the best overall model is MongoDB ST, managing to reduce time by more than $50\%$ compared to the worst model.
\end{enumerate}

\subsubsection{Global Queries} \label{ssec:global_queries}
  
Regarding global queries, we demonstrate the results for both hep-Th, hep-Ph and US Patents datasets. Our experiments include more query processing modes than local queries, since we distinguish between RA, RR and ID approaches. Recall that we focus on global queries in this paper, since this is the expected bottleneck for a vertex-centric approach. 

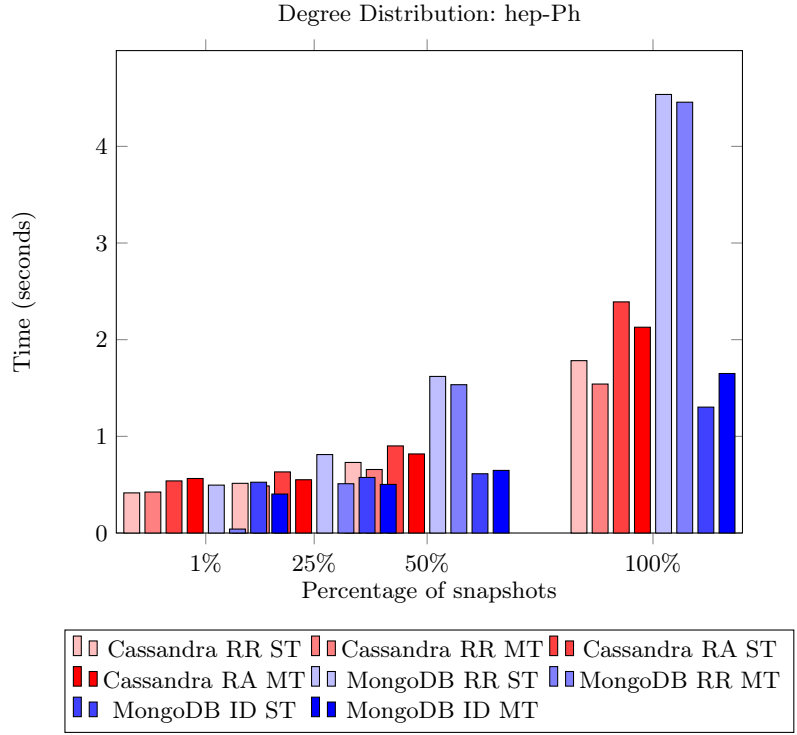
\begin{figure}[tb!]
    \centering
    \begin{tikzpicture}
    \begin{axis}[
        ybar,
        bar width=6pt,
        width=0.65\textwidth,
        height=8cm,
        xlabel={Percentage of snapshots},
        ylabel={Time (seconds)},
        xtick={1, 25, 50, 100},
        xticklabels={1\%, 25\%, 50\%, 100\%},
        ymin=0,
        enlarge x limits=0.2,
        legend style={at={(0.5,-0.2)}, anchor=north, legend columns=3},
        title={Degree Distribution: hep-Ph}
    ]
    \addplot[fill=red!25] coordinates {(1, 0.415) (25, 0.514) (50, 0.73) (100, 1.783)};
    \addplot[fill=red!50] coordinates {(1, 0.424) (25, 0.487) (50, 0.657) (100, 1.541)};
    \addplot[fill=red!75] coordinates {(1, 0.539) (25, 0.632) (50, 0.901) (100, 2.391)};
    \addplot[fill=red!100] coordinates {(1, 0.564) (25, 0.551) (50, 0.818) (100, 2.129)};
    \addplot[fill=blue!25] coordinates {(1, 0.495523) (25, 0.811547) (50, 1.619903) (100, 4.536927)};
    \addplot[fill=blue!50] coordinates {(1, 0.040665) (25, 0.509874) (50, 1.534313) (100, 4.456814)};
    \addplot[fill=blue!75] coordinates {(1, 0.525435) (25, 0.574647) (50, 0.612906) (100, 1.302343)};
    \addplot[fill=blue!100] coordinates {(1, 0.403122) (25, 0.502675) (50, 0.647746) (100, 1.650177)};
    
    \legend{Cassandra RR ST, Cassandra RR MT, Cassandra RA ST, Cassandra RA MT, MongoDB RR ST, MongoDB RR MT, MongoDB ID ST, MongoDB ID MT}
    \end{axis}
    \end{tikzpicture}
    \caption{Degree Distribution Comparison for the Hep-Ph dataset.}\label{lo::degdistrPh}
\end{figure}

\begin{figure}[tb!]
    \centering
    \begin{tikzpicture}
    \begin{axis}[
        ybar,
        bar width=6pt,
        width=0.65\textwidth,
        height=8cm,
        xlabel={Percentage of snapshots},
        ylabel={Time (seconds)},
        xtick={1, 25, 50, 100},
        xticklabels={1\%, 25\%, 50\%, 100\%},
        ymin=0,
        enlarge x limits=0.2,
        legend style={at={(0.5,-0.2)}, anchor=north, legend columns=3},
        title={Degree Distribution: hep-Th}
    ]
    \addplot[fill=red!25] coordinates {(1, 0.341) (25, 0.385) (50, 0.561) (100, 1.495)};
    \addplot[fill=red!50] coordinates {(1, 0.322) (25, 0.356) (50, 0.453) (100, 1.122)};
    \addplot[fill=red!75] coordinates {(1, 0.444) (25, 0.489) (50, 0.707) (100, 2.12)};
    \addplot[fill=red!100] coordinates {(1, 0.458) (25, 0.471) (50, 0.548) (100, 1.777)};
    \addplot[fill=blue!25] coordinates {(1, 0.433683552) (25, 0.643024999) (50, 1.83663717) (100, 3.545010769)};
    \addplot[fill=blue!50] coordinates {(1, 0.025516728) (25, 0.260315595) (50, 0.98099632) (100, 3.798652161)};
    \addplot[fill=blue!75] coordinates {(1, 0.474023645) (25, 0.482565655) (50, 0.477226309) (100, 0.823955787)};
    \addplot[fill=blue!100] coordinates {(1, 0.426538782) (25, 0.343976512) (50, 0.523912672) (100, 1.38256873)};
    
    \legend{Cassandra RR ST, Cassandra RR MT, Cassandra RA ST, Cassandra RA MT, MongoDB RR ST, MongoDB RR MT, MongoDB ID ST, MongoDB ID MT}
    \end{axis}
    \end{tikzpicture}
    \caption{Degree Distribution Comparison for the Hep-Th dataset.}\label{lo::degdistrDataset2}
\end{figure}
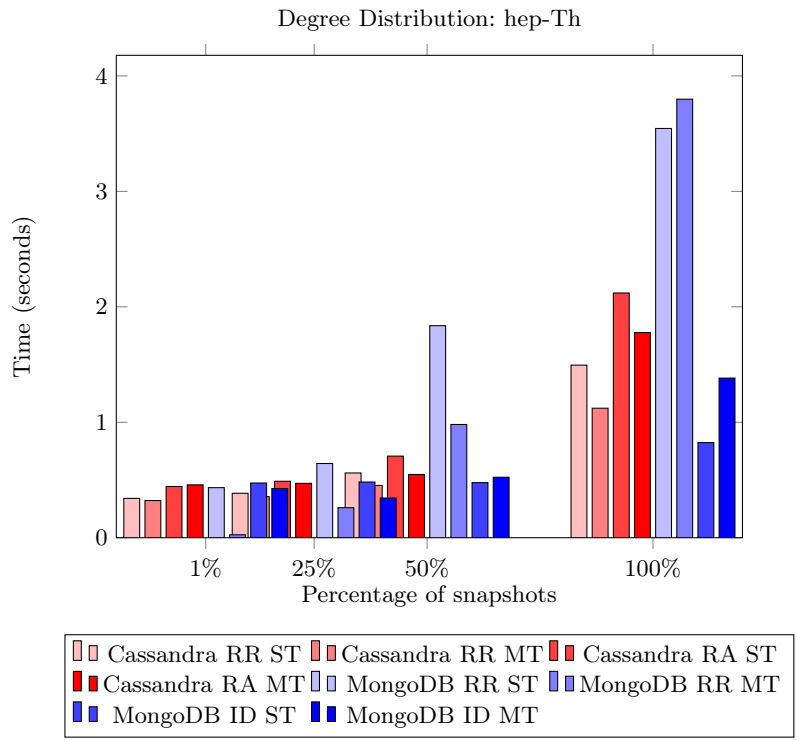

\begin{figure}[tb!]
    \centering
    \begin{tikzpicture}
    \begin{axis}[
        ybar,
        bar width=6pt,
        width=0.65\textwidth,
        height=8cm,
        xlabel={Percentage of snapshots},
        ylabel={Time (seconds)},
        xtick={1, 25, 50, 100},
        xticklabels={1\%, 25\%, 50\%, 100\%},
        ymin=0,
        enlarge x limits=0.2,
        legend style={at={(0.5,-0.2)}, anchor=north, legend columns=3},
        title={Degree Distribution: US Patents}
    ]
    \addplot[fill=red!25] coordinates {(1, 43.377) (25, 66.088) (50, 109.042) (100, 0)};
    \addplot[fill=red!50] coordinates {(1, 16.436) (25, 25.212) (50, 43.141) (100, 147.22)};
    \addplot[fill=red!75] coordinates {(1, 44.643) (25, 58.756) (50, 85.4) (100, 0)}; 
    \addplot[fill=red!100] coordinates {(1, 19.744) (25, 19.031) (50, 29.42) (100, 0)}; 
    \addplot[fill=blue!25] coordinates {(1, 105.8074) (25, 158.7280373) (50, 240.1218307) (100, 534.4064362)};
    \addplot[fill=blue!50] coordinates {(1, 96.60468205) (25, 145.2002344) (50, 190.9513053) (100, 460.9010082)};
    \addplot[fill=blue!75] coordinates {(1, 14.86375561) (25, 22.9919427) (50, 38.65604913) (100, 119.7668451)};
    \addplot[fill=blue!100] coordinates {(1, 10.03178729) (25, 12.51106349) (50, 17.20044147) (100, 226.8035888)};
    
    \legend{Cassandra RR ST, Cassandra RR MT, Cassandra RA ST, Cassandra RA MT, MongoDB RR ST, MongoDB RR MT, MongoDB ID ST, MongoDB ID MT}
    \end{axis}
    \end{tikzpicture}
    \caption{Degree Distribution Comparison for the US Patents dataset.}\label{lo::degdistrDataset3}
\end{figure}
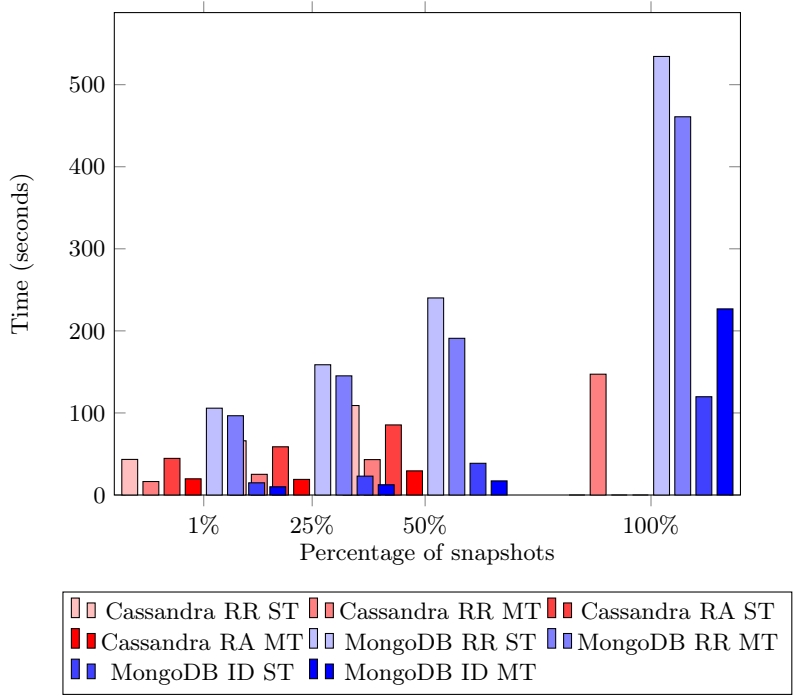

\textbf{Vertex Degree Distribution Query.} The summary results are shown in Figures \ref{lo::degdistrPh}, \ref{lo::degdistrDataset2}, and \ref{lo::degdistrDataset3}, from which we draw the following observations:
\begin{enumerate}
    \item Due to memory restrictions, among the Cassandra models only Cassandra ST RR was able to execute for $100\%$ of the snapshots in the US-Patents dataset. On the other hand, all MongoDB models managed to execute for all query percentages.
    
    \item The best models differ based on the size of the dataset and the percentage of snapshots used. Cassandra MT RR is the best Cassandra model for smaller datasets in all snapshots, while for the bigger Dataset (US-Patents) until $50\%$ of the snapshot Cassandra MT RA is the best Cassandra model, while for $100\%$ of the snapshots Cassandra MT RR is the best Cassandra model. For MongoDB, when $100\%$ of the snapshots are queried, the best model is MongoDB ST ID. In all cases, the best model is a MongoDB model, outperforming the best Cassandra model with a percentage difference between $3.17\% - 31.05\%$ in the smaller datasets, and $20.56\% - 52.42\%$ in US Patents.
    
    \item In most cases, the Cassandra MT models, outperform the corresponding Cassandra ST models, with a percentage difference up to $28.31\%$ in the smaller datasets, increasing to $102.14\%$ in US-Patents. 
    
    \item In MongoDB in most cases the MT model outperforms the corresponding ST model, by up to $85\%$ percentage difference, although when the query includes more than $50\%$ of the snapshots MongoDB ST ID outperforms the corresponding MT model by up to $76.82\%$ percentage difference. 
    
\end{enumerate}


Finally, while Cassandra requires less space to store the data since it builds fewer indexes and adopts a different storage approach, the MongoDB approach requires less memory on the client while executing the query. This is due to the iterative approach that was adopted in MongoDB, as well as due to the adoption of the ID query processing method. The space required for the three datasets in Cassandra ST was $31.0$ MB, $37.4$ MB and $1.83$ GB, and for the MT was $45.7$ MB, $55.5$ MB and $3.10$ GB, respectively. The space for MongoDB ST was $89.70$ MB, $107.37$ MB, and $4.84$ GB, and for MT was $218.34$ MB, $260.87$ MB and $10.96$ GB, respectively. On the other hand, MongoDB exhibited a speedup larger than $2\times$ when inserting data. 
In the RR and RA techniques both in MongoDB and Cassandra, we need to store either vid or the whole documents before processing them. On the other hand, clients using the ID approach need to retrieve only the document that is currently being processed, and as a result, the storage complexity depends on the execution of the query.

\begin{figure}[h!]
    \centering
    \begin{tikzpicture}
    \begin{axis}[
        ybar,
        bar width=6pt,
        width=0.65\textwidth,
        height=8cm,
        xlabel={Percentage of snapshots},
        ylabel={Time (seconds)},
        xtick={1, 25, 50, 100},
        xticklabels={1\%, 25\%, 50\%, 100\%},
        ymin=0,
        enlarge x limits=0.2,
        legend style={at={(0.5,-0.2)}, anchor=north, legend columns=3},
        title={Degree Distribution: hep-Ph}
    ]
    \addplot[fill=red!25] coordinates {(1, 4.122) (25, 4.985) (50, 5.955) (100, 8.472)};
    \addplot[fill=red!50] coordinates {(1, 0.428) (25, 0.757) (50, 1.408) (100, 3.478)};
    \addplot[fill=red!75] coordinates {(1, 6.86) (25, 7.164) (50, 7.74) (100, 10.125)};
    \addplot[fill=red!100] coordinates {(1, 4.534) (25, 4.615) (50, 5) (100, 7.546)};
    \addplot[fill=blue!25] coordinates {(1, 0.5105) (25, 3.1953) (50, 5.5647) (100, 33.1515)};
    \addplot[fill=blue!50] coordinates {(1, 0.2045) (25, 3.5547) (50, 12.5097) (100, 44.8042)};
    \addplot[fill=blue!75] coordinates {(1, 0.4994) (25, 0.5551) (50, 0.6426) (100, 1.0339)};
    \addplot[fill=blue!100] coordinates {(1, 1.5390) (25, 1.5116) (50, 1.8479) (100, 2.1)};
    
    \legend{Cassandra RR ST, Cassandra RR MT, Cassandra RA ST, Cassandra RA MT, MongoDB RR ST, MongoDB RR MT, MongoDB ID ST, MongoDB ID MT}
    \end{axis}
    \end{tikzpicture}
    \caption{(Cluster) Degree Distribution Comparison for the Hep-Ph dataset.}\label{cl::degdistrPh}
\end{figure}
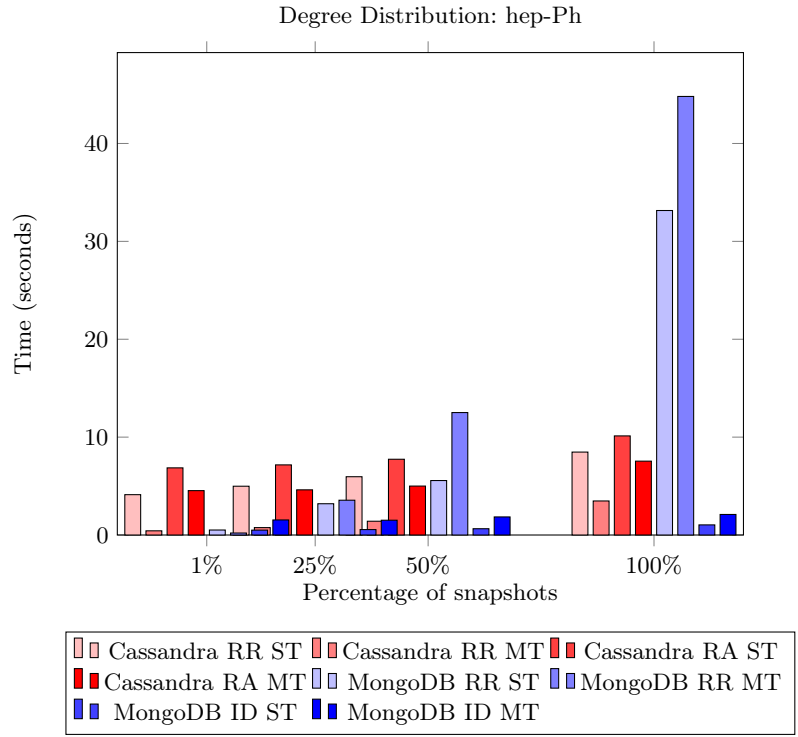

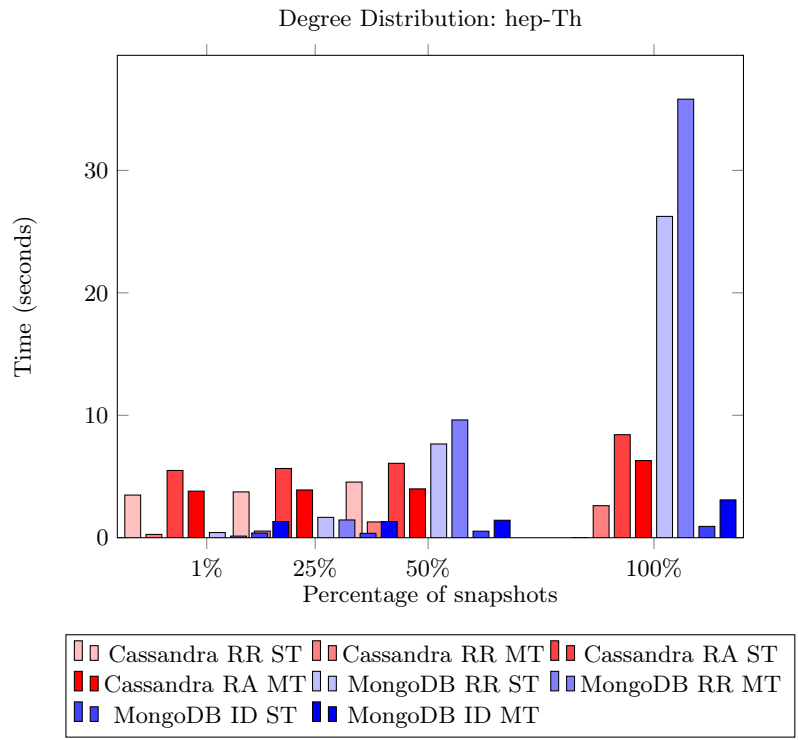
\begin{figure}[tb!]
    \centering
    \begin{tikzpicture}
    \begin{axis}[
        ybar,
        bar width=6pt,
        width=0.65\textwidth,
        height=8cm,
        xlabel={Percentage of snapshots},
        ylabel={Time (seconds)},
        xtick={1, 25, 50, 100},
        xticklabels={1\%, 25\%, 50\%, 100\%},
        ymin=0,
        enlarge x limits=0.2,
        legend style={at={(0.5,-0.2)}, anchor=north, legend columns=3},
        title={Degree Distribution: hep-Th}
    ]
    \addplot[fill=red!25] coordinates {(1, 3.485) (25, 3.743) (50, 4.546) (100, 0)};
    \addplot[fill=red!50] coordinates {(1, 0.271) (25, 0.544) (50, 1.291) (100, 2.617)};
    \addplot[fill=red!75] coordinates {(1, 5.494) (25, 5.656) (50, 6.079) (100, 8.415)};
    \addplot[fill=red!100] coordinates {(1, 3.805) (25, 3.902) (50, 3.99) (100, 6.3)};
    \addplot[fill=blue!25] coordinates {(1, 0.4224) (25, 1.6655) (50, 7.6563) (100, 26.2444)};
    \addplot[fill=blue!50] coordinates {(1, 0.1329) (25, 1.4522) (50, 9.6209) (100, 35.8173)};
    \addplot[fill=blue!75] coordinates {(1, 0.3842) (25, 0.3580) (50, 0.5360) (100, 0.9250)};
    \addplot[fill=blue!100] coordinates {(1, 1.3194) (25, 1.3158) (50, 1.4284) (100, 3.0894)};
    
    \legend{Cassandra RR ST, Cassandra RR MT, Cassandra RA ST, Cassandra RA MT, MongoDB RR ST, MongoDB RR MT, MongoDB ID ST, MongoDB ID MT}
    \end{axis}
    \end{tikzpicture}
    \caption{ (Cluster) Degree Distribution Comparison for the hep-Th dataset.}\label{cl::degdistrTh}
\end{figure}

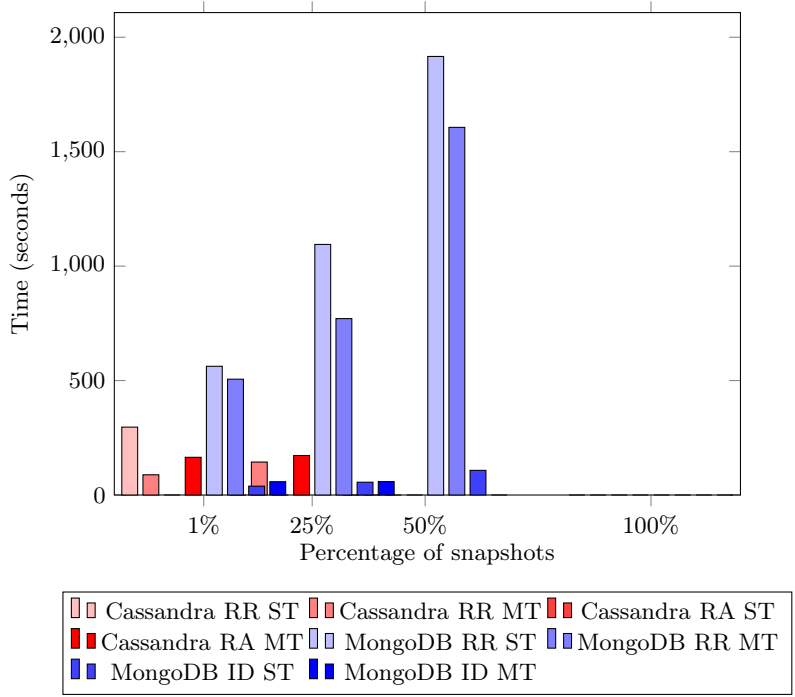
\begin{figure}[tb!]
    \centering
    \begin{tikzpicture}
    \begin{axis}[
        ybar,
        bar width=6pt,
        width=0.65\textwidth,
        height=8cm,
        xlabel={Percentage of snapshots},
        ylabel={Time (seconds)},
        xtick={1, 25, 50, 100},
        xticklabels={1\%, 25\%, 50\%, 100\%},
        ymin=0,
        enlarge x limits=0.2,
        legend style={at={(0.5,-0.2)}, anchor=north, legend columns=3},
        title={Degree Distribution: US Patents}
    ]
    \addplot[fill=red!25] coordinates {(1, 296.614) (25, 0) (50, 0) (100, 0)};
    \addplot[fill=red!50] coordinates {(1, 88.362) (25, 144.009) (50, 0) (100, 0)};
    \addplot[fill=red!75] coordinates {(1, 0) (25, 0) (50, 0) (100, 0)};
    \addplot[fill=red!100] coordinates {(1, 164.763) (25, 172.601) (50, 0) (100, 0)};
    \addplot[fill=blue!25] coordinates {(1, 562.4560) (25, 1094.8539) (50, 1916.1088) (100, 0)};
    \addplot[fill=blue!50] coordinates {(1, 506.0467) (25, 770.7150) (50, 1606.2976) (100, 0)};
    \addplot[fill=blue!75] coordinates {(1, 38.6754) (25, 55.9566) (50, 107.7702) (100, 0)};
    \addplot[fill=blue!100] coordinates {(1, 58.1414) (25, 58.6418) (50, 0) (100, 0)};
    
    \legend{Cassandra RR ST, Cassandra RR MT, Cassandra RA ST, Cassandra RA MT, MongoDB RR ST, MongoDB RR MT, MongoDB ID ST, MongoDB ID MT}
    \end{axis}
    \end{tikzpicture}
    \caption{(Cluster) Degree Distribution Comparison for the US Patents dataset.}\label{cl::degdistrus}
\end{figure}

Regarding global queries in the cluster environment, we demonstrate the results for the hep-Th, hep-Ph, and US Patents datasets. We employ only the Degree Distribution query, as we observed similar results with the Average Degree query. In addition, based on the previous experiments, we use only the most promising approaches. In Cassandra, we experiment with RA and RR, while in MongoDB we experiment with RR and ID both in ST and MT models.

The results can be seen in the Figures \ref{cl::degdistrPh}, \ref{cl::degdistrTh} and \ref{cl::degdistrus}. In the following, we summarize the main observations from these experiments. 

\begin{enumerate}
	\item In the two smaller datasets (hep-Ph, hep-Th), the MongoDB ST ID Model is the best-performing model, managing to reduce time from $30\%$ up to $64.6\%$ when compared to the best Cassandra model, with the exception when querying for only one snapshot. In this case (the query involves one snapshot), the best-performing model in the small datasets is Cassandra MT, reducing time by more than $10\%$ when compared to the best MongoDB model.
	
    \item In Cassandra, the MT models manage to outperform the corresponding ST models. The RR approach reduces execution time by $58.9\%$ to $89.6\%$, $71.6\%$ to $92.2\%$ and $70\%$ in the datasets hep-Ph, hep-Th, and US Patents respectively. Similarly, RA reduces time by $25.4\%$ to $35.5\%$ and $25.1\%$ to $34.3\%$ for the two smaller datasets.
    
    \item In MongoDB, the ST models manage to outperform the corresponding MT models. Recall that the ST model corresponds to a pure vertex-centric approach as in the case of HiNode. The RR approach reduces execution time by $10.1\%$ to $59.9\%$, $20\%$ to $26.7\%$, and $10\%$ to $29.6\%$ in the datasets hep-Ph, hep-Th and US Patents correspondingly. Similarly, ID reduces time by at most $70.2\%$ in hep-Ph, $62.4\%$ to $72.9\%$ in hep-Th, and $4\%$ to $33.4\%$ in US Patents.
	
	\item Due to low memory on the client side, some experiments on the US Patents dataset failed. This indicates which models and algorithms are less memory-demanding for the client and increase the distribution of tasks. Those are the MongoDB RR ST, MongoDB RR MT, and MongoDB ID ST that were able to execute the query for up to $50\%$ of the graph snapshots of the US Patents dataset.
	
	\end{enumerate}

\subsubsection{Transactions}

\begin{table}[tb!]
\begin{center}
\begin{tabular}{|c|c|l|l|l|}
\hline
 &
   &
  \multicolumn{1}{c|}{\textbf{hep-Th}} &
  \multicolumn{1}{c|}{\textbf{hep-Ph}} &
  \multicolumn{1}{c|}{\textbf{US Patents}} \\ \hline
\multirow{2}{*}{\textbf{MongoDB}}   & ST                            & 2.6 & 3.0 & 118.3  \\ \cline{2-5} 
                                    & MT                            & 1.96   & 2.4    & 105.27    \\ \hline
\multirow{2}{*}{\textbf{Cassandra}} & ST                            & 3.91  & 4.44 & 191.2 \\ \cline{2-5} 
                                    & MT                            & 3.72  & 4.44 & 188.2 \\ \hline
\end{tabular}
\caption{The database creation time from a sequence of transactions (insertions, deletions, modifications) for each model and each dataset in a local machine (in minutes).}
  \label{table:insertions}
\end{center}
\end{table}

\begin{table}[tb!]
\centering
\begin{tabular}{|c|c|c|c|c|}
\hline
\multicolumn{2}{|c|}{} & \textbf{hep-Th} & \textbf{hep-PH} & \textbf{US Patents} \\ \hline
\multirow{2}{*}{\textbf{MongoDB}}   & ST & 25  & 37.9  & 1318.8  \\ \cline{2-5} 
                                    & MT & 58.7  & 63.4  & 2879.4  \\ \hline
\multirow{2}{*}{\textbf{Cassandra}} & ST & 40.3  & 42.5  & 1747.5  \\ \cline{2-5} 
                                    & MT & 49.8  & 58.7  & 2485.1  \\ \hline
\end{tabular}
\caption{The database creation time from a sequence of transactions (insertions, deletions, modifications) for each model and each dataset in the cluster (in minutes).}
\label{table:clusterins}
\end{table}

Regarding testing transactions, our experiments involve the operations of insertion, modification, and deletion. These actions are performed at the beginning of our tests while creating the database. No tests were made with other OLAP queries (e.g., one hop-queries) that could be executed concurrently. In Table \ref{table:insertions}, we provide the total database construction time for each model in local mode (single machine). By observing the results we can reach the following deductions: 
\begin{enumerate}
    \item The ST models have comparable efficiency with the MT models, with less than $13\%$ time reduction when MT model is used.
    \item Comparing the best Cassandra with the best MongoDB models, the latter outperforms Cassandra by $79.1\%$ to $89.79\%$. 
\end{enumerate}

In Table \ref{table:clusterins} we report on the construction times in the cluster environment. We observe a rather different behavior compared to inserting the data locally, which may be attributed to many factors. The most important reasons for why the ST model outperforms the MT model is the replication of data in the cluster, the persistence of data throughout the cluster, as well as the use of the majority consistency level. In addition, since we used strict rules in the cluster in order to better support the consistency of the data, it is expected that we will have slower transaction times. 

\begin{enumerate}
    \item The ST models outperforms the MT models, by $118.43\% - 134\%$ in MongoDB and $23.8\% - 42.35\%$ in Cassandra.
    \item Comparing the best Cassandra with the best MongoDB models, the latter outperforms Cassandra by $32.54\% - 61.2\%$. 
\end{enumerate}

\subsection{Experiments on a Historical Graph Dataset with Time Intervals} \label{sec:historical}

To enhance the support for our experimental investigations, we sought to evaluate our models utilizing workloads designed for historical graph analysis. Upon an exhaustive examination of available resources in this domain, it became evident that there is a notable scarcity of openly accessible datasets containing historical information conducive to our research objectives. Consequently, we tried to come up with alternative solutions.

The Linked Data Benchmark Council (LDBC) \cite{angles2024ldbc}, renowned for its contributions to benchmarking workloads and dataset generation, has developed a suite of tools tailored for graph database evaluation. Among these tools, the ``LDBC SNB Datagen`` generator stands out for its provision of comprehensive information regarding data creation and deletion events. Although this dataset does not encompass updates on parameters, it aligns closely with our current dataset test case and enjoys widespread recognition within the research community.

While historical information pertinent to our research goals is absent from the interactive LDBC SNB workload, it is available within the dataset generated by the aforementioned data generator. To leverage this dataset as a testing workload, we needed to move to an event-based paradigm. The original format of the data, wherein all node and edge information, including parameter data and temporal details, are encapsulated within a single row, posed a challenge. To better emulate real-time operational scenarios, we restructured the dataset and ordered the events chronologically, primarily impacting the end time \footnote{This transformation process is publicly available in \url{https://github.com/alexspitalas/HGDataset}.}.

\subsubsection{The Schema of the Historical Graph Dataset}
\begin{figure}[tb!]
  \centering
  \includegraphics[width=\columnwidth]{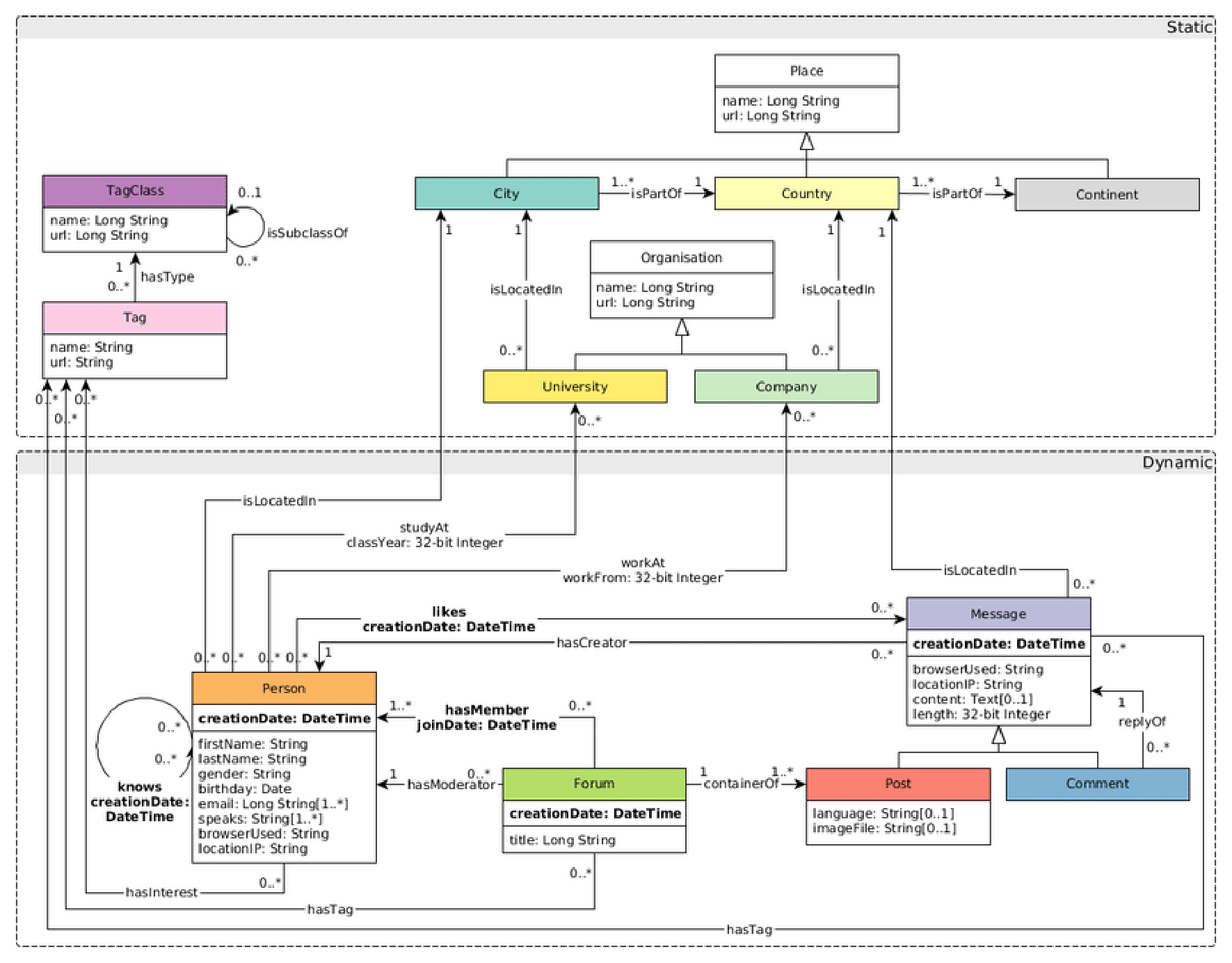}
  \caption{The general LDBC SNB Schema.}
  \label{fig:LDBC-Schema}
\end{figure}

The dataset generated from the Linked Data Benchmark Council (LDBC) for Social Network Benchmark (SNB) exhibits diverse forms contingent upon user configurations. Users possess the capability to adjust cardinality numbers, as well as the minimum and maximum number of values straightforwardly. Furthermore, with a more intricate endeavor, users can manipulate additional parameters. The default schema of SNB, depicted in Figure~\ref{fig:LDBC-Schema}, contains a static part and a dynamic part. The static graph encompasses entities that remain invariant throughout the benchmark's duration, while dynamic entities are subject to modification or removal during the benchmark execution.

\begin{figure}[tb!]
  \centering
  \includegraphics[]{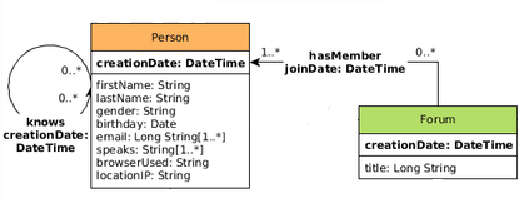}
  \caption{The reduced LDBC SNB Schema used for our testing cases.}
  \label{fig:LDBC-Schema-Cropped}
\end{figure}

The LDBC SNB offers a standardized approach for evaluating and benchmarking database systems. A key aspect of this benchmarking process involves determining scale factors (SFs) to accommodate systems of varying sizes and resource limitations.
SFs are calculated based on the ASCII size of resulting output files, measured in Gibibytes (GiB), utilizing a specific serializer as outlined in Section 3.4.2 of the benchmark documentation. These SFs represent different dataset sizes: SF1 equals 1 GiB, SF100 equals 100 GiB, and SF10,000 equals 10,000 GiB. 

By default, all SFs span a three-year period beginning in 2010, and their calculation is based on scaling the number of individuals within the network. This means that the benchmark data generated at different SFs reflects various sizes of social networks, with SF10,000 representing a significantly larger and more intricate network compared to SF1.
More information can be found in \cite{ldbc-specs}

In the initial phase of our proof-of-concept benchmark, we aimed to simplify the graph structure, focusing solely on dynamic entities. Consequently, we employed a reduced schema, as illustrated in Figure~\ref{fig:LDBC-Schema-Cropped}, comprising only of two entities: Person and Forum, along with the corresponding edges ``Person Knows Person`` and ``Forum has Member Person``. Additionally, both creationDate and deletionDate attributes are included for entities.
We used the default parameters in the configuration with SF3 and SF10 sizes that are included in our reduced dataset, with only ``Person`` entities and ``Person Knows Person`` edges in ``SF3`` and ``SF10`` datasets. We also used one more extended dataset ``SF3 Person and Forum`` that also includes the data about ``Forum`` entity and ``Forum has Member Person`` edges. The dataset statistics are shown in Table~\ref{tab:SFdatasets}.

\begin{table}[tb!]
\begin{center}
\begin{tabular}{|l|c|c|}
\hline
 Name & \# of vertices & \# of edges \\ \hline
 SF3 & $25870$ & 668430  \\ \hline
 SF10 & 60800 & 2304951  \\ \hline
 SF3 Person and Forum & 285499 & 10499492  \\ \hline
 \end{tabular}
\caption{Details about the datasets used.}
\label{tab:SFdatasets}
\end{center}
\end{table}


\subsubsection{Streaming Experiments}

In a streaming environment, data typically arrives in the form of events rather than batches. Therefore, the approach used in the previous experiments that relied on node or edge loading in batches would not be suitable for this scenario. 
To address this limitation, we designed an event-based approach for both Cassandra and MongoDB implementations. In the context of an event-based system, the transactions represent usually simple events with minimal changes instead of making multiple data changes.
For our experiments, we used the LDBC dataset generated using SF3 and SF10 with some modifications.

To establish an event-based system, we first need to identify the fundamental transactions required to represent a historical graph comprehensively. We divided the transactions into two categories: node/edge/parameter insertion and deletion. It is important to note that deletion only sets the ending time of an object since nothing is deleted in a historical graph unless explicitly stated. The insertion process requires only the start time of the object and sets the end time to infinity. The HiNode model is used to store all objects. The API was modified to accommodate the event-based approach, and some index modifications were made to enhance performance. 
More specifically, the following instructions are supported: 
\begin{itemize}
    \item insert\_node($start$): It creates a new empty historical node $v$ with a valid interval $[start, +\infty)$, meaning that the node is valid for all time instants after $start$. The same holds for all other operations as well, with respect to the use of the special symbol $+ \infty$.
    \item insert\_edge($u, v, start$): It creates a new edge with a valid interval $[start, +\infty)$, which must be contained in the valid interval of both nodes $u$ and $v$. The same check (although not mentioned) holds for the rest of the operations. 
    \item insert\_property($p, f, val, start$): It creates a new property $f$ in the historical object $p$ (node or edge) with value $val$ and a valid interval $[start, +\infty)$.
    \item delete\_node($v, end$): Node $v$ has its valid interval shrunk since it is invalidated in the time range $(end, +\infty)$. This means that all properties of $v$ and edges adjacent to $v$ must be checked so that their valid intervals are still legitimate. In case they are not, then their valid intervals are shrunk. 
    \item delete\_edge($e, end$): Edge $e$ has its valid interval shrunk, since it is invalidated in the time range $(end,+\infty)$. This means that all properties of $e$ must be checked so that their valid intervals are still legitimate. In case they are not, then their valid intervals are shrunk as well. 
    \item delete\_property($p, f, end$): The valid interval of property $f$ of the historical object $p$ is shrunk, since it is invalidated in the time range $(end, +\infty)$. 
\end{itemize}
Using combinations of these operations, we can support historical graph streaming applications.

Regarding space efficiency, the observations are the same with Sec. \ref{ssec:global_queries} with the only difference being that the SF3 Person and Forum MongoDB MT implementation, may take up to $7\times$ more space than the corresponding Cassandra implementation. This is happening due to the preallocation of memory in MongoDB documents. In addition, one can see that there is a disagreement between Cassandra and MongoDB related to ST and MT models. While Cassandra MT uses less space that ST, for MongodDB it holds the inverse. The reason is that Cassandra is a wide-column store, while MongoDB is a document database. The former is best suited to the case where we split the columns of a table into multiple smaller ones like we did in the MT model. The latter is more efficient when the documents store more data rather than having many documents with small data each. Thus, especially in the SF datasets were we have more attributes, the latter has better results in the ST model compared to MT.

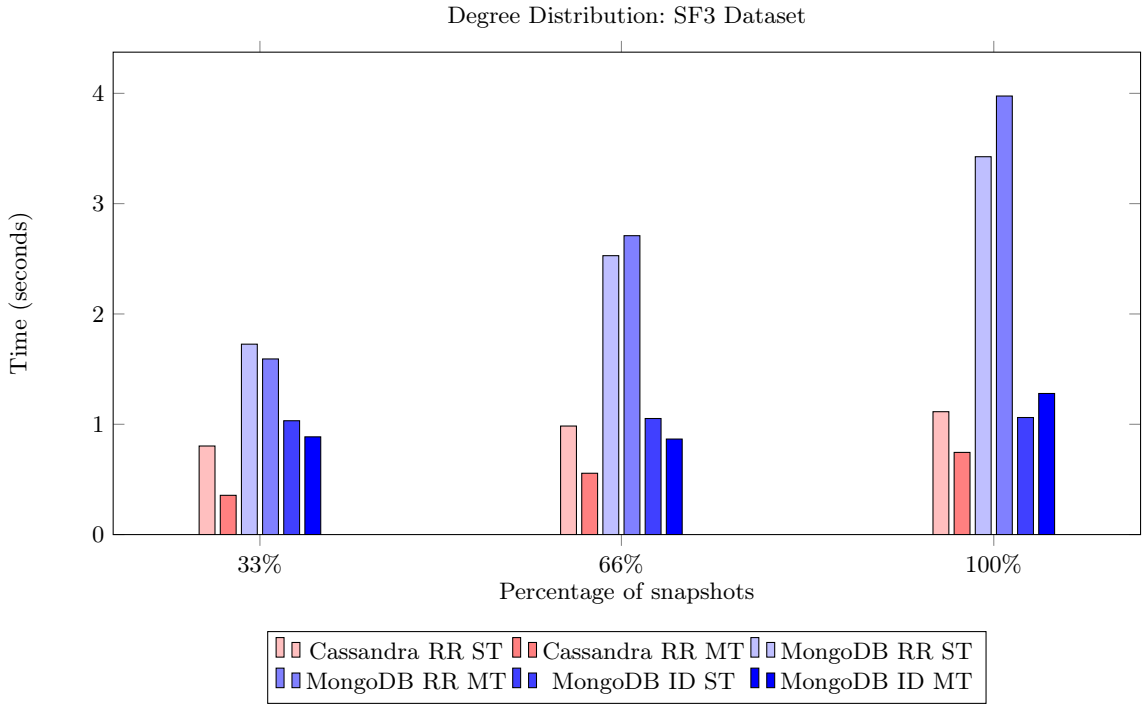
\begin{figure}[h!]
    \centering
    \begin{tikzpicture}
    \begin{axis}[
        ybar,
        bar width=6pt,
        width=\textwidth,
        height=8cm,
        xlabel={Percentage of snapshots},
        ylabel={Time (seconds)},
        xtick={33, 66, 100},
        xticklabels={33\%, 66\%, 100\%},
        ymin=0,
        enlarge x limits=0.2,
        legend style={at={(0.5,-0.2)}, anchor=north, legend columns=3},
        title={Degree Distribution: SF3 Dataset}
    ]
    \addplot[fill=red!25] coordinates {(33, 0.803) (66, 0.984) (100, 1.114)};
    \addplot[fill=red!50] coordinates {(33, 0.356) (66, 0.556) (100, 0.745)};
    \addplot[fill=blue!25] coordinates {(33, 1.726213186) (66, 2.528346246) (100, 3.425452434)};
    \addplot[fill=blue!50] coordinates {(33, 1.592201849) (66, 2.709619439) (100, 3.97565386)};
    \addplot[fill=blue!75] coordinates {(33, 1.032059287) (66, 1.052297131) (100, 1.061275231)};
    \addplot[fill=blue!100] coordinates {(33, 0.886065696) (66, 0.866056966) (100, 1.279145272)};
    
    \legend{Cassandra RR ST, Cassandra RR MT,
            MongoDB RR ST, MongoDB RR MT, MongoDB ID ST, MongoDB ID MT}
    \end{axis}
    \end{tikzpicture}
    \caption{Results for the vertex degree distribution query on the SF3 dataset in a single machine.}\label{lo::degdistrSF3}
\end{figure}

\begin{figure}[h!]
    \centering
    \begin{tikzpicture}
    \begin{axis}[
        ybar,
        bar width=6pt,
        width=\textwidth,
        height=8cm,
        xlabel={Percentage of snapshots},
        ylabel={Time (seconds)},
        xtick={33, 66, 100},
        xticklabels={33\%, 66\%, 100\%},
        ymin=0,
        enlarge x limits=0.2,
        legend style={at={(0.5,-0.2)}, anchor=north, legend columns=3},
        title={Degree Distribution: SF10 Dataset}
    ]
    \addplot[fill=red!25] coordinates {(33, 3.081) (66, 3.606) (100, 4.201)};
    \addplot[fill=red!50] coordinates {(33, 1.254) (66, 1.965) (100, 2.424)};
    \addplot[fill=blue!25] coordinates {(33, 3.88976726) (66, 7.040520568) (100, 8.283051256)};
    \addplot[fill=blue!50] coordinates {(33, 4.933377771) (66, 8.796623712) (100, 11.55528453)};
    \addplot[fill=blue!75] coordinates {(33, 2.253477432) (66, 2.393002063) (100, 2.256636678)};
    \addplot[fill=blue!100] coordinates {(33, 2.440369762) (66, 2.87849296) (100, 3.331482038)};
    
    \legend{Cassandra RR ST, Cassandra RR MT,
            MongoDB RR ST, MongoDB RR MT,
            MongoDB ID ST, MongoDB ID MT}
    \end{axis}
    \end{tikzpicture}
    \caption{Results for the vertex degree distribution query on the SF10 dataset in a single machine.}\label{lo::degdistrSF10}
\end{figure}
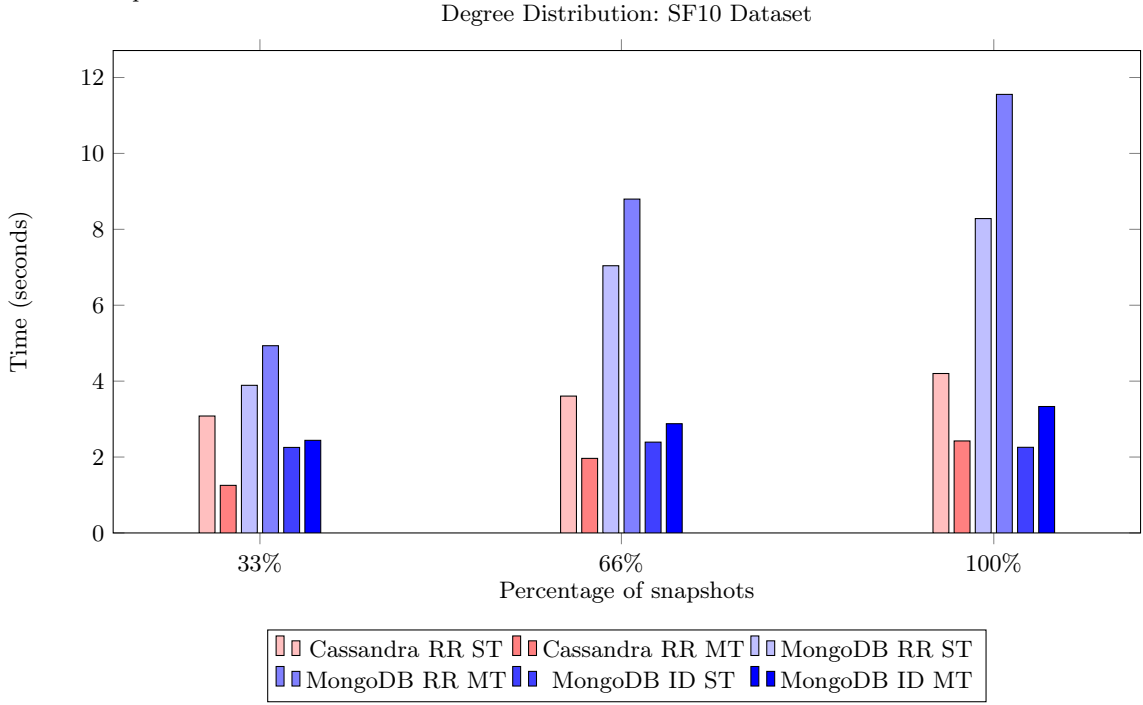

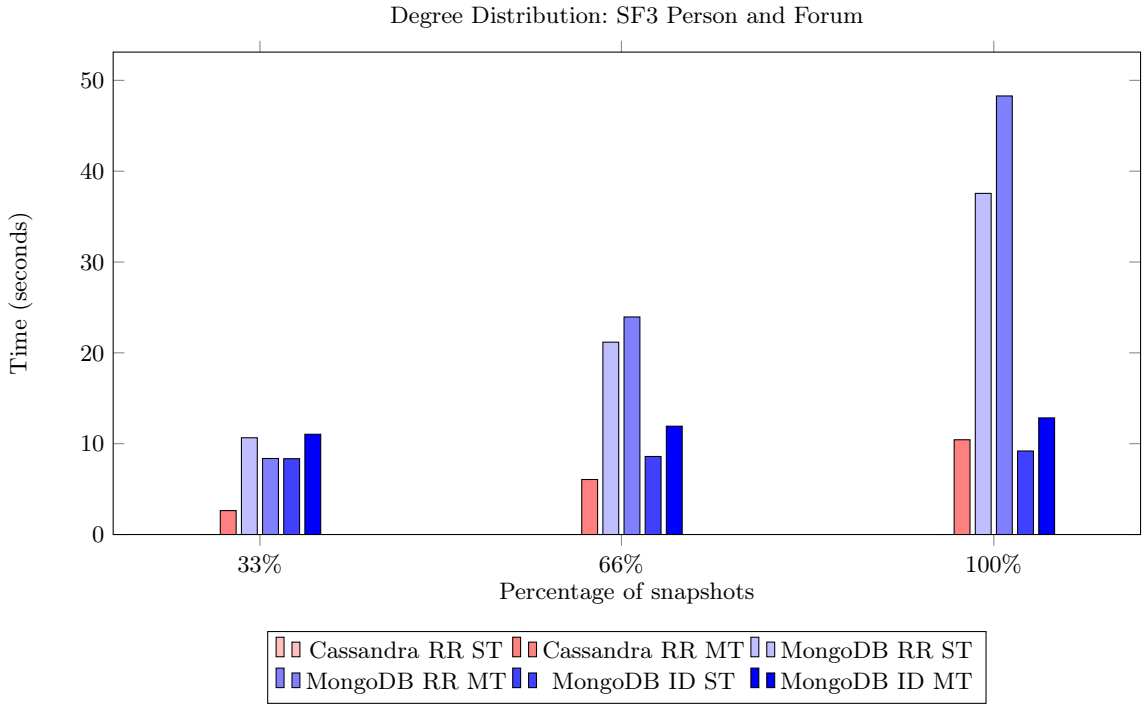
\begin{figure}[h!]
    \centering
    \begin{tikzpicture}
    \begin{axis}[
        ybar,
        bar width=6pt,
        width=\textwidth,
        height=8cm,
        xlabel={Percentage of snapshots},
        ylabel={Time (seconds)},
        xtick={33, 66, 100},
        xticklabels={33\%, 66\%, 100\%},
        ymin=0,
        enlarge x limits=0.2,
        legend style={at={(0.5,-0.2)}, anchor=north, legend columns=3},
        title={Degree Distribution: SF3 Person and Forum}
    ]
    \addplot[fill=red!25] coordinates {(33, 0) (66, 0) (100, 0)};
    \addplot[fill=red!50] coordinates {(33, 2.636) (66, 6.06) (100, 10.434)};
    \addplot[fill=blue!25] coordinates {(33, 10.65320841) (66, 21.17739889) (100, 37.55557766)};
    \addplot[fill=blue!50] coordinates {(33, 8.371887709) (66, 23.95036227) (100, 48.27963573)};
    \addplot[fill=blue!75] coordinates {(33, 8.349725686) (66, 8.594419126) (100, 9.198364163)};
    \addplot[fill=blue!100] coordinates {(33, 11.03899838) (66, 11.92776017) (100, 12.84086506)};
    
    \legend{Cassandra RR ST, Cassandra RR MT,
            MongoDB RR ST, MongoDB RR MT,
            MongoDB ID ST, MongoDB ID MT}
    \end{axis}
    \end{tikzpicture}
    \caption{Results for the vertex degree distribution query on the SF3 Person and Forum dataset in a single machine.}\label{lo::degdistrSF3PF}
\end{figure}

\begin{table}[tb!]
\begin{center}
\begin{tabular}{|c|c|l|l|l|}
\hline
 &
   &
  \multicolumn{1}{c|}{\textbf{SF3}} &
  \multicolumn{1}{c|}{\textbf{SF10}} &
  \multicolumn{1}{c|}{\textbf{SF3 Forum}} \\ \hline
\multirow{2}{*}{\textbf{MongoDB }}   & ST                            & 7.20  & 25.75 & 106.22  \\ \cline{2-5} 
                                    & MT                            & 1.72   & 5.93     & 21.458   \\ \hline
\multirow{2}{*}{\textbf{Cassandra}} & ST                            & 10.30  & 36.65 & $+\infty$ \\ \cline{2-5} 
                                    & MT                            & 1.745 & 5.762 & 20.331 \\ \cline{2-5} 
                                 \hline
\end{tabular}
\caption{The time to create the database from a sequence of insertions for each model and each dataset in a single machine (in minutes - $+\infty$ indicates that the construction did not complete). }
  \label{table:insertionsNew}
\end{center}
\end{table}

\begin{table}[]
\begin{center}
\begin{tabular}{|c|c|l|l|l|}
\hline
 &
   &
  \multicolumn{1}{c|}{\textbf{SF3}} &
  \multicolumn{1}{c|}{\textbf{SF10}} &
  \multicolumn{1}{c|}{\textbf{SF3 Extended}} \\ \hline
\multirow{2}{*}{\textbf{MongoDB}}   & ST                            & 132.7 MB  & 437.6 MB & 1.9GB  \\ \cline{2-5} 
                                    & MT                            & 145.3 MB   & 479.6MB    & 2.22 GB    \\ \hline
 \multirow{2}{*}{\textbf{CASSANDRA}} & ST                            & 62.5 MB  & 278 MB  & N/A \\ \cline{2-5} 
                                    & MT                            & 20 MB  & 108 MB & 336.7 MB \\ 
                                 \hline
\end{tabular}
\caption{Size of the database for each model and each dataset respectively.}
  \label{table:size}
\end{center}
\end{table}

After inserting the data into each model (Cassandra/MongoDB, ST/MT), we tested both local and global queries. In particular, we employed the one-hop and the degree distribution query using the same test environment as in the previous experiments. For one-hop queries, some vertices are randomly selected and we compute the average of the queries, while for degree distribution, we query for 3 time spans, for one, two and three years (that is, for $33.3\%$, $66.6\%$ and $100\%$ of the graph respectively). The degree distribution is computed per year, meaning that when we calculate the degree distribution in the time range of three years, the result will include the degree distribution for the first, second and third year separately. The models we choose to test for the MongoDB are the best ones based on the preceding experiments. The results of the experimentation in a single machine can be summarized in the following observations:
\begin{enumerate}
    \item In Table~\ref{table:insertionsNew}, the time to build the historical graph database by consecutive insertions for various models is depicted. We can observe that the MT models have the best performance with a speedup of up to 7.2x compared to the corresponding ST models. MT models have comparable performance in Cassandra and MongoDB, but ST models have the best performance in MongoDB, outperforming the corresponding Cassandra model by up to $34.9\%$.

    \item In Figures~\ref{lo::degdistrSF3}, \ref{lo::degdistrSF10}, and \ref{lo::degdistrSF3PF}, the time needed to execute a degree distribution query is depicted for various models. For Cassandra, the best model is RR MT, which consistently outperforms its ST counterpart across all datasets. In general, Cassandra shows significant performance improvements with MT models, compared to ST models, with RR MT being up to 2.4x faster than RR ST in some cases  Conversely, in MongoDB, ST models generally outperform MT models, especially when using more percentage of snapshots and when larger datasets are used. Especially, the ID ST model performs best, showing stable performance across different snapshot percentages.
    
    \item Cassandra MT emerges as the superior model, showcasing significant performance advantages over MongoDB in most cases. For the SF3 dataset, Cassandra MT RR outperforms MongoDB ST ID by 29.80\% - 65.51\% , translating to a speedup of up to 3.1×. These performance gains are consistent across both SF3 and SF10 datasets, in the later dataset Cassandra MT RR improves the performance by up to 44.35\% in 33\% and 66\% while when we query for 100\% of the snapshots MongoDB ST ID is the best-performing model improving the performance by 7.42\% compared to the best Cassandra model. Notably, the ID processing query mode in MongoDB demonstrates exceptional stability, with execution time increases limited to at most 15\% as the query percentage of snapshots grows. This contrasts sharply with other modes and models, particularly Cassandra, where time increases can reach up to 100\% in the SF3 Person and Forum dataset as snapshot percentages increase.
  
    \item The MongoDB ST ID model is becoming more efficient compared to other models as the size of the dataset increases. The size of the datasets can be seen in Table~\ref{tab:SFdatasets}, although the size needed in its database to store them is depicted in Table~\ref{table:size}. Moving from SF3 to SF10 we increased more than $3.5$ times the number of edges while we can observe for Cassandra ST and MT models a $3.7$ and $3.4$ times increase of time spent in the query, while for MongoDB the time increase is 2.2 and 3 times for ST ID and MT ID models respectively, showing better scale while data increases. Between SF10 and SF3 Person and Forum, the edges are increased $4.5$ times, while the query time increased $3.7$ times for the MongoDB ST ID model, $4.14$ for MongoDB MT ID model, and 2-4 times for Cassandra MT model.

    \item Cassandra has an advantage because it exhibits high-throughput transactions, but because of the nature of the schema, it will get setbacks when multiple updates happen in the parameters or edges (which are not considered in the current test cases).
\end{enumerate}

\begin{table}[h!]
\centering
\begin{tabular}{|c|c|c|c|c|}
\hline
\multicolumn{2}{|c|}{} & \textbf{SF3} & \textbf{SF10} & \textbf{Extended SF3} \\ \hline
\multirow{2}{*}{\textbf{Cassandra}} & ST & 103.3  & 373.4  & N/A \\ \cline{2-5} 
                                    & MT & 22.2  & 72.6  &  270.1  \\ \hline
\multirow{2}{*}{\textbf{MongoDB}}   & ST & 251.3  & 906.4  & 2703.9  \\ \cline{2-5} 
                                    & MT & 143.2  & 268.8  & 1055.1  \\ \hline
\end{tabular}
\caption{Construction time (a sequence of insertions) comparison for SF3, SF10, and the extended SF3 datasets in the cluster (in minutes - N/A indicates that the construction was not carried out for this dataset for implementation reasons).}
\label{table:insertionsNewCluster}
\end{table}

\begin{figure}[h!]
    \centering
    \begin{tikzpicture}
    \begin{axis}[
        ybar,
        bar width=12pt,
        width=\textwidth,
        height=8cm,
        xlabel={Percentage of time used in the query},
        ylabel={Time (Seconds)},
        xtick={33, 66, 100},
        xticklabels={33\%, 66\%, 100\%},
        ymin=0,
        enlarge x limits=0.2,
        legend style={at={(0.5,-0.2)}, anchor=north, legend columns=2},
        title={Degree Distribution Comparison: SF3}
    ]
    \addplot[fill=red!50] coordinates {(33, 6.091) (66, 7.375) (100, 8.546)};
    \addplot[fill=red!100] coordinates {(33, 1.65) (66, 2.551) (100, 2.948)};
    \addplot[fill=blue!25] coordinates {(33, 9.0261) (66, 13.2069) (100, 17.0203)};
    \addplot[fill=blue!50] coordinates {(33, 16.4673) (66, 31.3941) (100, 41.7185)};
    \addplot[fill=blue!75] coordinates {(33, 2.3039) (66, 2.3654) (100, 2.4753)};
    \addplot[fill=blue!100] coordinates {(33, 3.5878) (66, 3.5374) (100, 3.5176)};
    
    \legend{Cassandra RR ST, Cassandra RR MT, MongoDB RR ST, MongoDB RR MT, MongoDB ID ST, MongoDB ID MT}
    \end{axis}
    \end{tikzpicture}
    \caption{Degree distribution query comparison for the SF3 dataset in the cluster.}
    \label{cl:degDistrsf3}
\end{figure}
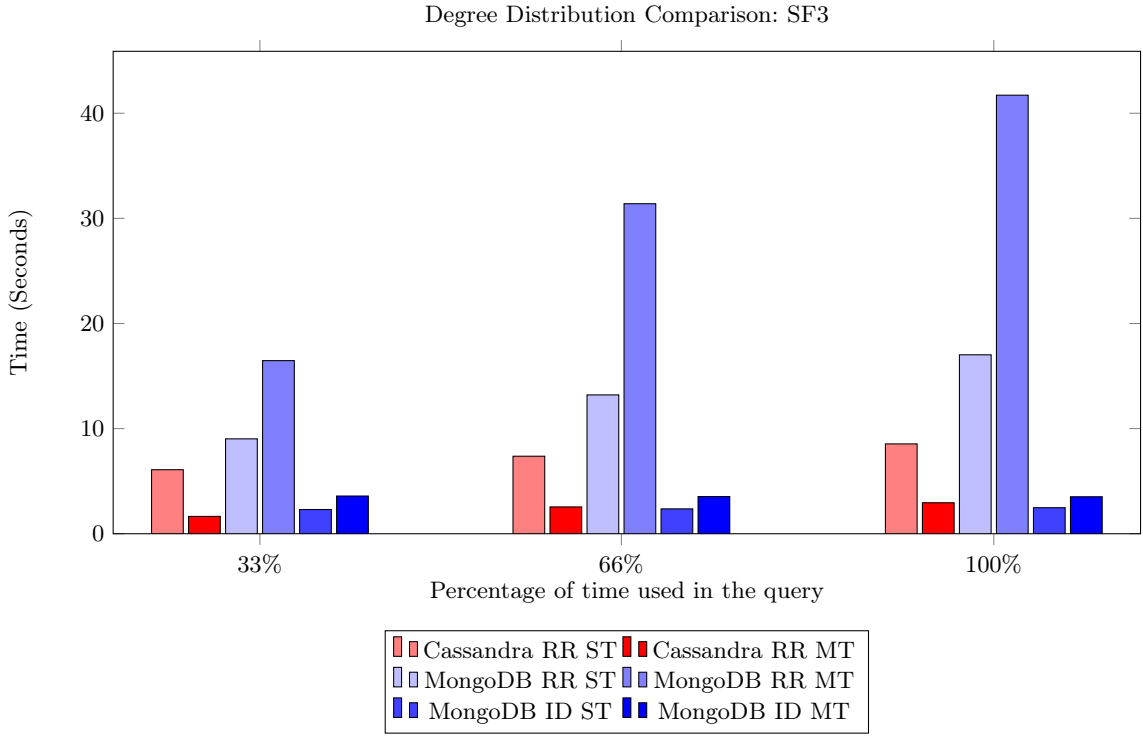

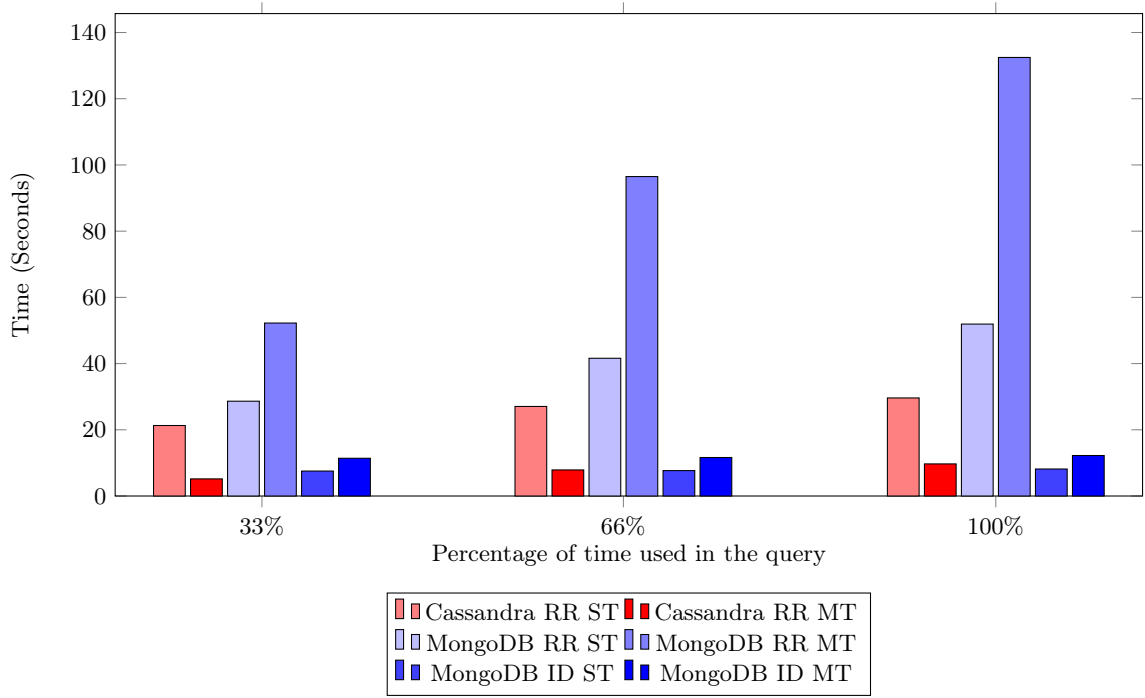
\begin{figure}[h!]
    \centering
    \begin{tikzpicture}
    \begin{axis}[
        ybar,
        bar width=12pt,
        width=\textwidth,
        height=8cm,
        xlabel={Percentage of time used in the query},
        ylabel={Time (Seconds)},
        xtick={33, 66, 100},
        xticklabels={33\%, 66\%, 100\%},
        ymin=0,
        enlarge x limits=0.2,
        legend style={at={(0.5,-0.2)}, anchor=north, legend columns=2},
        title={Degree Distribution Comparison: SF10}
    ]
    \addplot[fill=red!50] coordinates {(33, 21.296) (66, 27.052) (100, 29.621)};
    \addplot[fill=red!100] coordinates {(33, 5.171) (66, 7.875) (100, 9.698)};
    \addplot[fill=blue!25] coordinates {(33, 28.6389) (66, 41.6065) (100, 51.9328)};
    \addplot[fill=blue!50] coordinates {(33, 52.2474) (66, 96.4923) (100, 132.4731)};
    \addplot[fill=blue!75] coordinates {(33, 7.5376) (66, 7.6786) (100, 8.1584)};
    \addplot[fill=blue!100] coordinates {(33, 11.4019) (66, 11.6321) (100, 12.2352)};
    
    \legend{Cassandra RR ST, Cassandra RR MT, MongoDB RR ST, MongoDB RR MT, MongoDB ID ST, MongoDB ID MT}
    \end{axis}
    \end{tikzpicture}
    \caption{Degree distribution query comparison for the SF10 in the cluster.}
    \label{cl:degDistrsf10}
\end{figure}

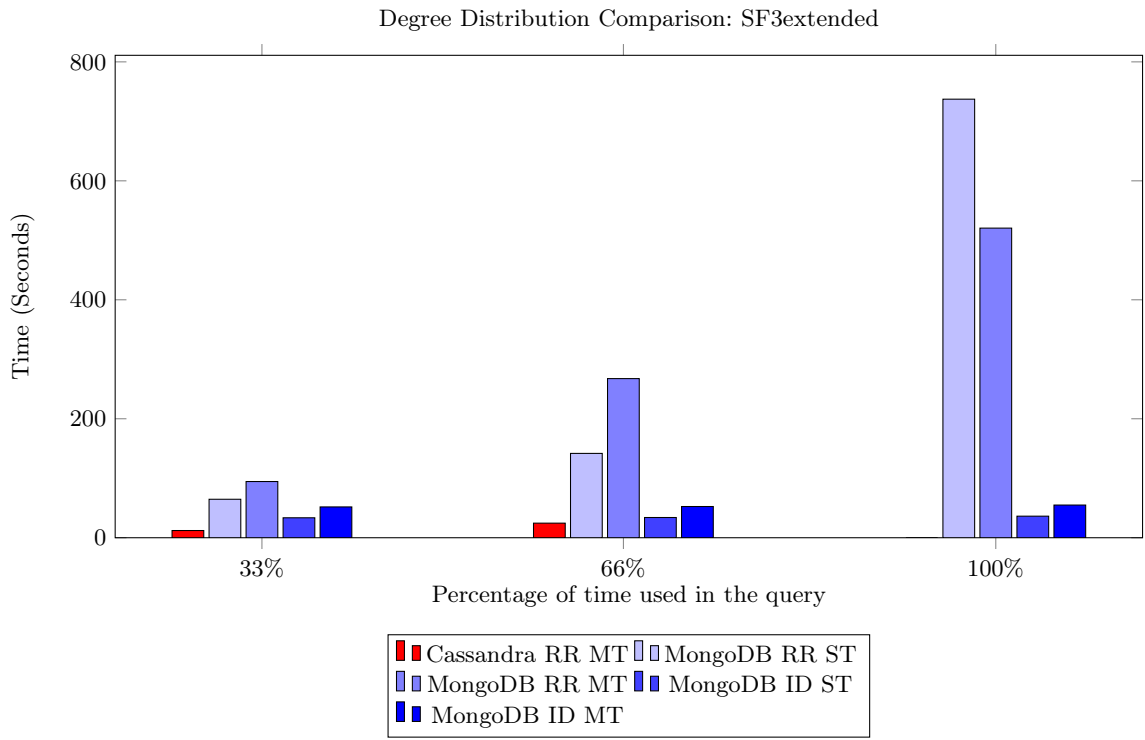
\begin{figure}[h!]
    \centering
    \begin{tikzpicture}
    \begin{axis}[
        ybar,
        bar width=12pt,
        width=\textwidth,
        height=8cm,
        xlabel={Percentage of time used in the query},
        ylabel={Time (Seconds)},
        xtick={33, 66, 100},
        xticklabels={33\%, 66\%, 100\%},
        ymin=0,
        enlarge x limits=0.2,
        legend style={at={(0.5,-0.2)}, anchor=north, legend columns=2},
        title={Degree Distribution Comparison: SF3extended}
    ]
    \addplot[fill=red!100] coordinates{(33, 12.132) (66, 24.532) (100, 0)};
    \addplot[fill=blue!25] coordinates{(33, 64.6904) (66, 141.7851) (100, 737.2989)};
    \addplot[fill=blue!50]  coordinates{(33, 94.5013) (66, 267.5368) (100, 520.5262)};
    \addplot[fill=blue!75]  coordinates{(33, 33.5425) (66, 34.0207) (100, 36.2666)};
    \addplot[fill=blue!100]  coordinates{(33, 51.8078) (66, 52.4947) (100, 54.9613)};
    
    \legend{Cassandra RR MT, MongoDB RR ST, MongoDB RR MT, MongoDB ID ST, MongoDB ID MT}
    \end{axis}
    \end{tikzpicture}
    \caption{Degree distribution query comparison for the extended SF3 dataset in the cluster.}
    \label{cl:degDistrsf3extended}
\end{figure}

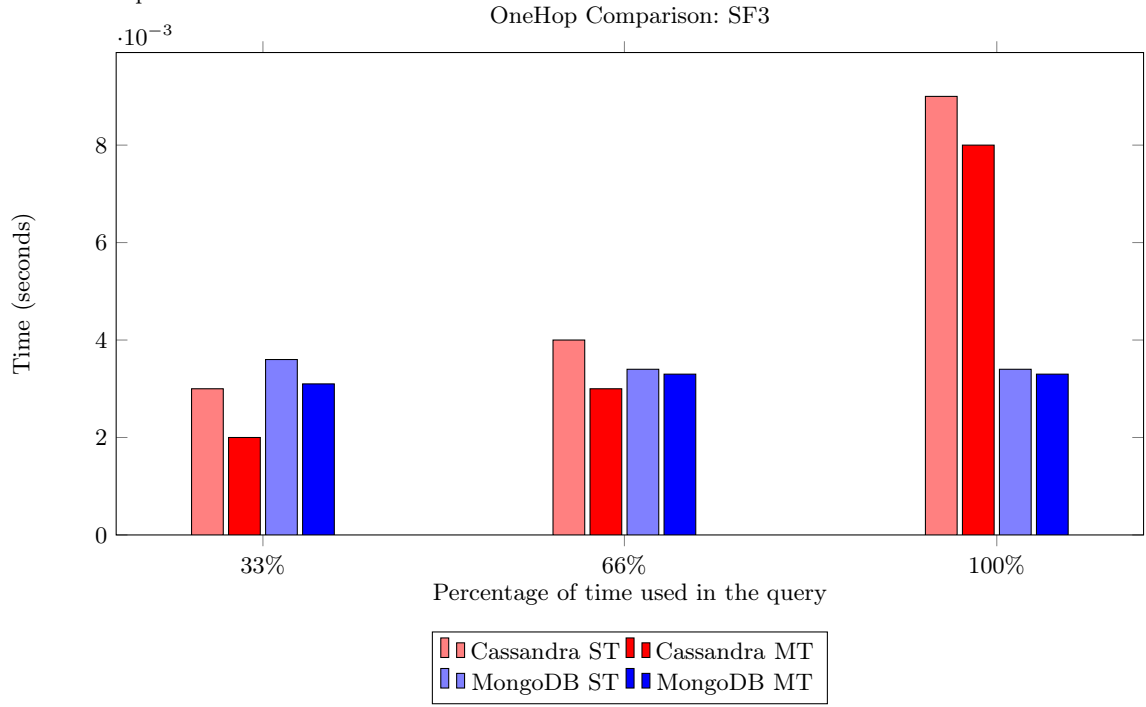
\begin{figure}[h!]
    \centering
    \begin{tikzpicture}
    \begin{axis}[
        ybar,
        bar width=12pt,
        width=\textwidth,
        height=8cm,
        xlabel={Percentage of time used in the query},
        ylabel={Time (seconds)},
        xtick={33, 66, 100},
        xticklabels={33\%, 66\%, 100\%},
        ymin=0,
        enlarge x limits=0.2,
        legend style={at={(0.5,-0.2)}, anchor=north, legend columns=2},
        title={OneHop Comparison: SF3}
    ]
    \addplot[fill=red!50] coordinates {(33, 0.003) (66, 0.004) (100, 0.009)};
    \addplot[fill=red!100] coordinates {(33, 0.002) (66, 0.003) (100, 0.008)};
    \addplot[fill=blue!50] coordinates {(33, 0.0036) (66, 0.0034) (100, 0.0034)};
    \addplot[fill=blue!100] coordinates {(33, 0.0031) (66, 0.0033) (100, 0.0033)};
    
    \legend{Cassandra ST, Cassandra MT, MongoDB ST, MongoDB MT}
    \end{axis}
    \end{tikzpicture}
    \caption{OneHop query comparison for the SF3 dataset in the cluster.}
    \label{cl:oneHopsf3}
\end{figure}

\begin{figure}[h!]
    \centering
    \begin{tikzpicture}
    \begin{axis}[
        ybar,
        bar width=12pt,
        width=\textwidth,
        height=8cm,
        xlabel={Percentage of time used in the query},
        ylabel={Time (seconds)},
        xtick={33, 66, 100},
        xticklabels={33\%, 66\%, 100\%},
        ymin=0,
        enlarge x limits=0.2,
        legend style={at={(0.5,-0.2)}, anchor=north, legend columns=2},
        title={OneHop Comparison: SF10}
    ]
    \addplot[fill=red!50] coordinates {(33, 0.004) (66, 0.004) (100, 0.005)};
    \addplot[fill=red!100] coordinates {(33, 0.002) (66, 0.003) (100, 0.004)};
    \addplot[fill=blue!50] coordinates {(33, 0.0035) (66, 0.0035) (100, 0.0034)};
    \addplot[fill=blue!100] coordinates {(33, 0.0032) (66, 0.0032) (100, 0.0032)};
    
    \legend{Cassandra ST, Cassandra MT, MongoDB ST, MongoDB MT}
    \end{axis}
    \end{tikzpicture}
    \caption{OneHop query comparison for the SF10 dataset in the cluster.}
    \label{cl:oneHopsf10}
\end{figure}
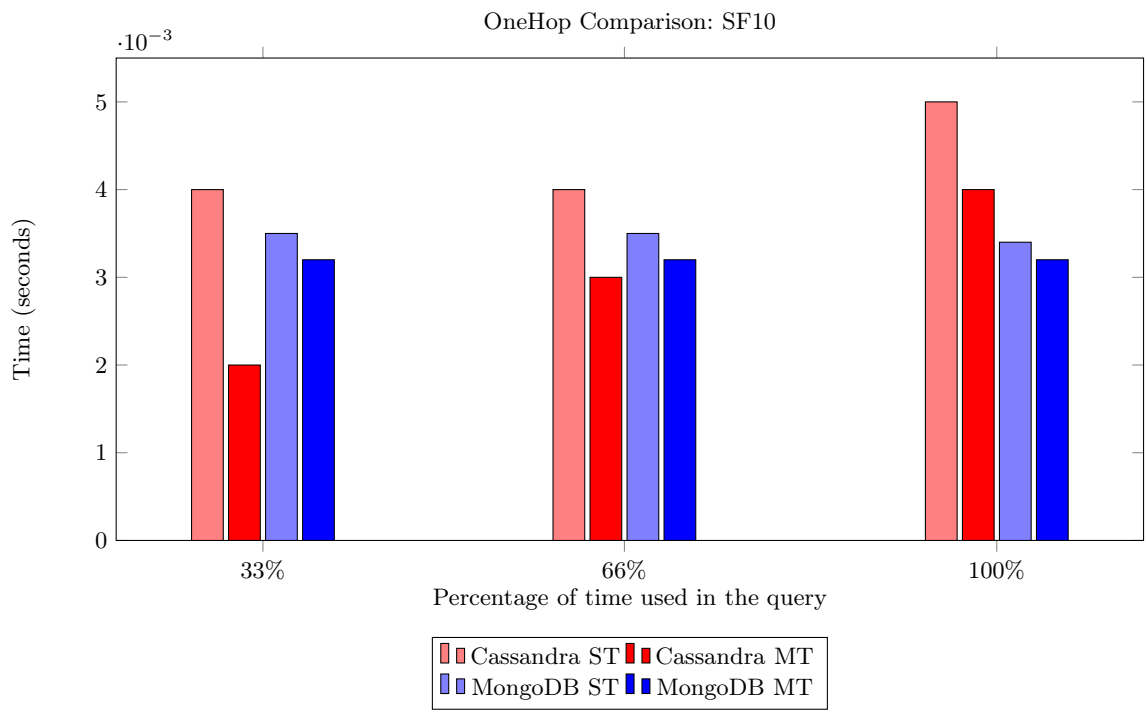

\begin{figure}[h!]
    \centering
    \begin{tikzpicture}
    \begin{axis}[
        ybar,
        bar width=12pt,
        width=\textwidth,
        height=8cm,
        xlabel={Percentage of time used in the query},
        ylabel={Time (seconds)},
        xtick={33, 66, 100},
        xticklabels={33\%, 66\%, 100\%},
        ymin=0,
        enlarge x limits=0.2,
        legend style={at={(0.5,-0.2)}, anchor=north, legend columns=2},
        title={OneHop Comparison: SF3extended}
    ]
    \addplot[fill=red!100] coordinates {(33, 0.002) (66, 0.002) (100, 0.002)};
    \addplot[fill=blue!50] coordinates {(33, 0.0045) (66, 0.0045) (100, 0.0047)};
    \addplot[fill=blue!100] coordinates {(33, 0.0036) (66, 0.0037) (100, 0.0040)};
    
    \legend{Cassandra MT, MongoDB ST, MongoDB MT}
    \end{axis}
    \end{tikzpicture}
    \caption{OneHop query comparison for the extended SF3 dataset in the cluster.}
    \label{cl:oneHopsf3extended}
\end{figure}
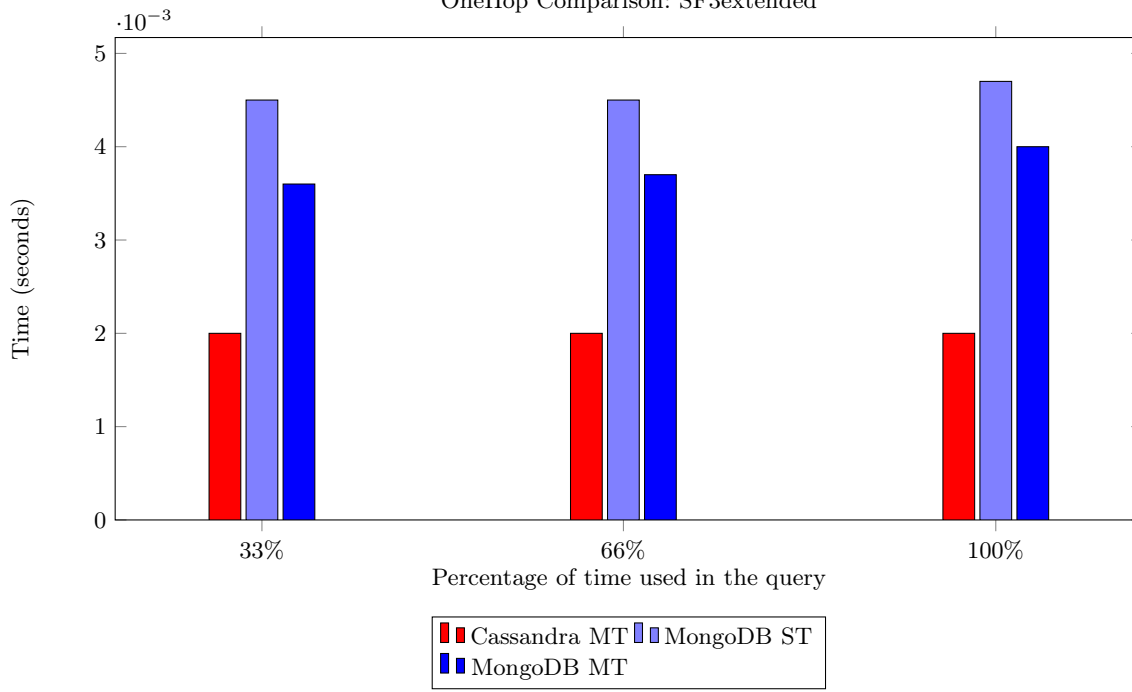

We executed both MongoDB and Cassandra models in the cluster and the results for Degree Distribution Query can be seen in Figures \ref{cl:degDistrsf3}, \ref{cl:degDistrsf10} and \ref{cl:degDistrsf3extended}, while for one hop query can be seen in Figures \ref{cl:oneHopsf3}, \ref{cl:oneHopsf10} and \ref{cl:oneHopsf3extended}. Also in Table~\ref{table:insertionsNewCluster}, is depicted the times to insert the data into the cluster. The observations, with some exceptions, are similar to the snapshot-based experiments:

\begin{enumerate}
    \item For Cassandra, the best-performing models are MT. In terms of degree distribution, they manage to reduce execution time by $65.4\%$ to $75.7\%$, while for one-hop queries, the reduction ranges from $11\%$ to $50\%$.
 
    \item For MongoDB, the best-performing models differ based on the application. For the global query degree distribution, ID ST is the best-performing model, managing to reduce execution time by $29.6\%$ up to $35.7\%$. However, for the case of one-hop local query, the best performing model is MT, reducing the execution time by $2.5\%$ up to $20.7\%$.

    \item Comparing the best Cassandra and MongoDB models, the dominant model differs on the application, size of the dataset or query percentage. 
    
    \item On the one-hop query, except when running for $100\%$ of the graph in SF3 and SF10, the best-performing model is Cassandra MT, managing to reduce execution time by $5.5\%$ up to $45.8\%$ compared to the best-performing MongoDB model. At $100\%$ of the graph, MongoDB ST is the dominant model, managing to reduce the execution time by $58\%$ in SF3 and $19.9\%$ in SF10 when compared to the best Cassandra model. 
    
    \item Regarding the degree distribution query, in SF3 and SF10, Cassandra MT is the best-performing model while querying for $33\%$ of the graph, reducing the execution time by $39.6\%$ and $45\%$ compared to the best MongoDB model. For degree distribution queries that involve $66\%$ or $100\%$ of the graph, MongoDB ST ID is the best-performing model, managing to reduce the execution time by up to $16\%$. In the extended SF3 dataset, the best-performing model is Cassandra MT (when it can be executed) managing to reduce the execution time between $27.8\%$ up to $63.8\%$ compared to the best MongoDB alternative. It can be observed that while doing so, the percentage is being reduced while we query a larger percentage of time. 

\end{enumerate}

\subsection{Discussion}
Regarding the differences between Cassandra and MongoDB implementations we observe the following:
\begin{enumerate}
 \item \textit{Local Queries in Snapshots}: Cassandra is more efficient (but not by far) than MongoDB except when a large number of time instants is involved in the query. In this case, MongoDB is more efficient.
 \item \textit{Global Queries in Snapshots}: In cluster mode, MongoDB is superior to Cassandra when the ID approach is adopted. In the other cases and in local mode the results are mixed.
 \item \textit{Transactions}: MongoDB using ST is faster than Cassandra both in local and in cluster mode.
 \item \textit{Streaming}: Both in local and cluster mode, Cassandra and MongoDB have similar performance. For global queries, MongoDB with ID seems to be the best choice, while for local queries the results are mixed.
 \item In general, ST is best performing in MongoDB, while  MT in Cassandra.
\end{enumerate}

Regarding the comparison of Single Table (ST) model that corresponds to a pure-vertex centric model, and the Multiple Table (MT) model in which a diachronic node has been split into a small number of collections, we reach the following conclusions related to MongoDB:

\begin{enumerate}
 \item \textit{Local Queries in Snapshots:} In a cluster environment, our small set of local queries shows that ST is slightly better than MT. 
 \item \textit{Global Queries in Snapshots:} Both in a local machine and in a cluster environment, ST and MT have similar performance, perhaps with a slight improvement from the ST side in the cluster environment for the ID approach.
 \item \textit{Transactions and Streaming}: The ST model seems to cope better in a cluster environment (this is a difference when compared to the local machine environment) when compared to MT for transactions.
 \item \textit{Streaming Queries}: Looking at the experiments on the cluster, the ST model seems to be doing a little better in global queries while in local queries the MT model seems to be doing better.  
 \item ST has slightly better space usage than MT. However, this depends largely on how fine is the splitting to multiple tables for the MT model, and the difference can become quite larger for finer (more tables) splitting.
\end{enumerate}

All in all, there is no clear winner - and we did not expect for one - since the performance depends on the type of query, on the size of the query time interval, as well as on the dataset. However, the goal of this paper was to show that the vertex-centric approach in a NoSQL environment has certain merits even in the case of global queries. One can extend this work among several axes. One can look at different queries and come up with mixed workloads (OLTP and OLAP) of various types. In addition, a side-effect that we discovered during our research, is the need for a generator for historical graph workloads. Our great aspiration is to make an existing graph database incorporate the notion of time as a first-class citizen. The current research was targeted towards this goal regarding the storage model and a certain type of queries.

\section{Conclusion}
\label{sec:open}

In this work, we have shown how the HiNode vertex-centric approach to storing time-varying graphs can be implemented in MongoDB, so that significant improvements in global queries can be achieved compared to the previous NoSQL-based implementation (over $4\times$ in some cases), while we also implemented the streaming functionality in both our systems. We focused mainly on global queries, since this is the main expected bottleneck of a vertex-centric approach. We made extensive comparisons between Cassanda and MongoDB implementations of different flavors, highlighting in the process some differences when considering a local machine environment or a cluster environment. In the process, the need arose to create appropriate datasets and set the stage for a future extension of known generators for historical graphs and their workloads.

We test successfully the Hinode model (a vertex-centric model) in various cases including streaming scenarios, transactions, demanding a partial vertex-centric approach (MT) and comparing it with a full vertex centric approach (ST), snapshots and time-interval based representations. In addition, we tested the Hinode model in the case where the schema of the graph can change with the addition of new properties in the nodes/edges (by using the extended SF3 dataset).

\bibliographystyle{splncs04}
\bibliography{bibliography}

\end{document}